\documentclass[aps,twocolumn,english,superscriptaddress,amsmath,floatfix,longbibliography,floats,prl]{revtex4-1}
\usepackage[colorlinks=true,urlcolor=blue,citecolor=blue,linkcolor=blue]{hyperref}

\usepackage[T1]{fontenc}
\usepackage[utf8]{inputenc}
\usepackage{amssymb}
\usepackage{graphicx}
\usepackage{overpic} 
\usepackage{amsmath,color}
\usepackage{mathrsfs}
\usepackage{float}
\usepackage{amsmath,bm}

\usepackage{indentfirst}
\usepackage{txfonts}
\usepackage[normalem]{ulem}
\usepackage[table]{xcolor}
\usepackage{array,ragged2e}
\usepackage{grffile}
\usepackage{multirow}
\usepackage{algpseudocode}
\usepackage{epstopdf}

\usepackage[caption=false]{subfig}

\usepackage{graphicx,xcolor} 

\usepackage{comment}

\newcommand{\bra}[1]{\langle#1|}
\newcommand{\ket}[1]{|#1\rangle}
\newcommand{\norm}[1]{\langle#1|#1\rangle}

\begin{document}

	\title{Locality-Induced Hierarchical Backflow Wavefunctions for Correlated Fermions}
	
	\author{Yu-Tong Zhou}
	\affiliation{Anhui Province Key Laboratory of Quantum Network, University of Science and Technology of China, Hefei 230026, China}
	\affiliation{Synergetic Innovation Center of Quantum Information and Quantum Physics, University of Science and Technology of China, Hefei 230026, China}
	\affiliation{Hefei National Laboratory, University of Science and Technology of China, Hefei 230088, China}	
	\author{Zheng-Wei Zhou}
	\email{zwzhou@ustc.edu.cn}
	\affiliation{Anhui Province Key Laboratory of Quantum Network, University of Science and Technology of China, Hefei 230026, China}
	\affiliation{Synergetic Innovation Center of Quantum Information and Quantum Physics, University of Science and Technology of China, Hefei 230026, China}
	\affiliation{Hefei National Laboratory, University of Science and Technology of China, Hefei 230088, China}	
	\affiliation{Anhui Center for Fundamental Sciences in Theoretical Physics, University of Science and Technology of China, Hefei 230026, China}
	\author{Wen-Yuan Liu}
	\email{wyliu@zju.edu.cn}
	\affiliation{Institute for Advanced Study in Physics and School of Physics, Zhejiang University, Hangzhou 310027, China}
	
	\begin{abstract}

We show that locality provides a natural principle to hierarchically organize backflow wavefunctions. This leads us to propose a family of variational fermionic states, termed hierarchical backflow (HB) wavefunctions. The expressive power of HB is systematically improvable, controlled by a path depth $K$ which reflects the range of backflow correlations. At half-filling, the HB with $K=1$ already achieves high energy precision, with an accuracy around $0.5\%$ for system sizes from $4\times 4$ to $10\times 10$. At hole doping $n_h=0.125$, the method scales efficiently to $12\times16$ and $16\times16$ systems, and the energy systematically achieves higher accuracy with $K$ increasing, yielding a clear stripe phase. The HB further enables a local-nonlocal decomposition, naturally bridging to neural quantum states, while featuring compact representations and efficient optimization. Our work reveals locality as a natural organizing principle of backflow wavefunctions, opening a new framework with systematic improvability and interpretability for large-scale simulations of correlated fermion systems.

	\end{abstract}
	
	\maketitle
	
	\textit{Introduction.} Simulating and representing strongly correlated fermionic systems is a central task in condensed matter physics. 
    These systems host many extraordinary phenomena such as high-temperature superconductivity, Mott transitions, quantum magnetism~\cite{doi:10.1098/rspa.1963.0204,doi:10.1142/S0217979292000414,PhysRevB.70.035114,RevModPhys.66.763,PhysRevMaterials.3.054605,doi:10.1126/science.aam7127,xuCoexistenceSuperconductivityPartially2024,annurev:/content/journals/10.1146/annurev-conmatphys-031620-102024},  yet accurately characterizing their nature presents a fundamental challenge due to the exponentially large Hilbert space. To overcome this difficulty, a variety of powerful approaches have been developed, such as dynamical mean-field theory~\cite{DMFT1,DMFT2,cDMFT}, quantum Monte Carlo~\cite{CPAFQMC1,CPAFQMC2,DDMC,DiaMC,GFQMC,sorella2023} and tensor network methods~\cite{dmrg1992,verstraete2008,jiang2019,gong2021robust,li2023tangent,qu2024phase,corboz2010simulation,corboz2011,liu2017,liu2021,liu2022gapless,liu2022emergence,liu2025hubbard}, each offering distinct insights and capabilities. Nevertheless, due to this complexity, developing efficient and systematically improvable methods remains an urgent need.

    In recent years, neural network quantum states have provided a promising approach to simulate correlated fermion systems.
     Most such states are built upon backflow wavefunctions~\cite{luo2019backflow,fermionNN2022,cassellaDiscoveringQuantumPhase2023,liFermionicNeuralNetwork2022,cassellaNeuralNetworkVariational2024,clark2024unifying,zhou2024solving,PhysRevX.14.021030,Liang2026}, introduced by Feynman and Cohen in 1956 to describe liquid helium~\cite{feynmanEnergySpectrumExcitations1956}. A backflow wavefunction encodes the fermion sign structure by taking the form of Slater determinants, but with each single-particle orbital depending on the configuration of all other particles, a property known as backflow correlations. This allows it to capture many-body physics beyond the mean-field level while preserving fermionic antisymmetry. Yet despite its 70-year history, a clear organizing principle underlying the expressive power of backflow wavefunctions remains lacking. Existing approaches, though expressive, typically treat the configuration dependence of the orbitals as generic global functions, posing challenges for systematic interpretability, improvement, and optimization.

	\begin{figure}
		\centering
		\begin{minipage}[b]{0.32\columnwidth}
			\centering
			\begin{overpic}[width=1\columnwidth]{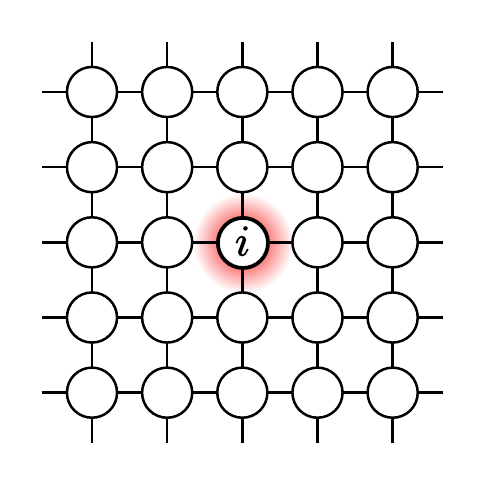}
				\put(5, 95){(a) $K=0$}  
			\end{overpic}
		\end{minipage}
		\begin{minipage}[b]{0.32\columnwidth}
			\centering
			\begin{overpic}[width=1\columnwidth]{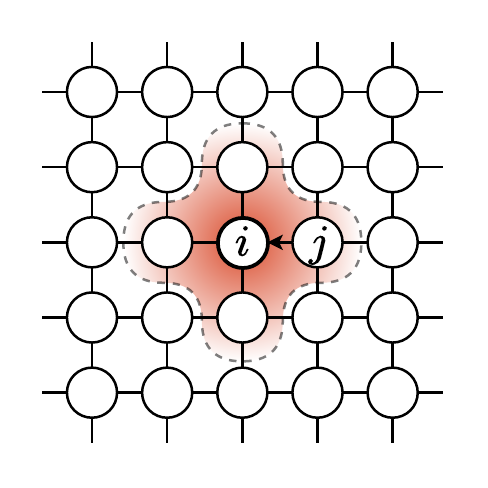}
				\put(5, 95){(b) $K=1$ }  
			\end{overpic}
		\end{minipage}
		\begin{minipage}[b]{0.32\columnwidth}
			\centering
			\begin{overpic}[width=1\columnwidth]{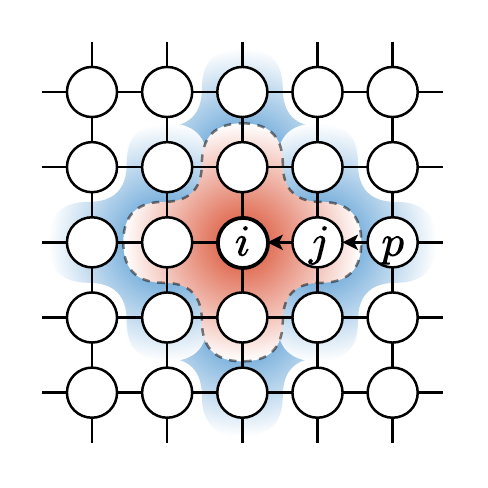}
				\put(5, 95){(c) $K=2$}  
			\end{overpic}
		\end{minipage}
        \hfill
		\begin{minipage}[b]{0.99\columnwidth}
			\centering
			\begin{overpic}[width=1\columnwidth]{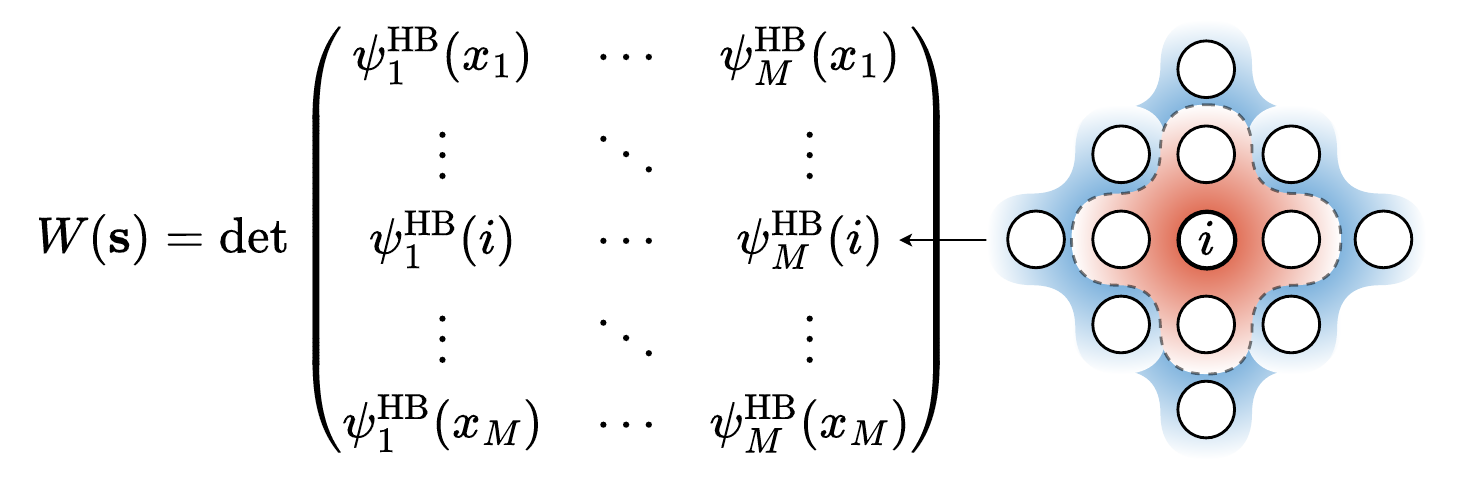}
				\put(2, 31){(d) }  
			\end{overpic}
		\end{minipage}
		\caption{Orbital structure of hierarchical backflow (HB) wavefunctions at a given site $i$ for different path depths $K$, illustrating how $K$ controls the range of backflow correlations. (a) $K=0$: the orbital depends only on site $i$, corresponding to the Hartree-Fock state with no backflow correlations. (b) $K=1$: the orbital at site $i$ depends on its nearby sites $j$. (c) $K=2$: the orbital further includes sites $p$ that are neighbors of $j$; increasing $K$ systematically enhances the expressive power. (d) Although each orbital depends only on local paths (here shown for $K=2$), the Slater determinant couples them to generate many-body correlations in the wavefunction amplitude $W(\mathbf{s})$.}
		\label{fig:localbackflow}
	\end{figure}

In this work, we reveal that locality provides a natural organizing principle that builds backflow correlations from iterated local transitions, rather than viewing it as a mysterious nonlocal feature. This perspective leads to a new family of variational fermionic states, which we term hierarchical backflow (HB) wavefunctions. Their expressive power is controlled by a path depth $K$, reflecting the locality-induced hierarchical structure of backflow correlations. For $K=0$, the ansatz reduces to the Hartree-Fock state, while increasing $K$ systematically improves the expressive power. Furthermore, the hierarchical backflow naturally leads to a local-nonlocal decomposition that bridges directly to neural network quantum states, enabling compact representations compared to conventional global-network architecture.

We demonstrate this framework on the two-dimensional Hubbard model. At half-filling, even with the smallest nontrivial depth $K=1$, the HB already achieves high energy precision, with a relative energy error around $0.5\%$ for system sizes from $4\times 4$ to $10\times 10$. At hole doping $n=0.125$, scaling to $12\times 16$ and $16\times 16$, the energy  progressively improves to higher accuracy with increasing $K$, exhibiting a clear stripe phase. Our work thus establishes locality as a natural organizing principle of backflow wavefunctions, opening a new systematically improvable and interpretable framework for large-scale simulations of strongly correlated fermion systems.

	\textit{Locality-induced hierarchy in backflows.} Consider a fermionic Hubbard-type Hamiltonian with nearest-neighbor interactions on an $N$-site lattice: 
	\begin{equation}
		H = \sum_{\langle ij \rangle} t_{ij}c_i^\dagger c_j + \sum_{\langle ik \rangle} U_{ik} n_i n_k,
	\end{equation}
	where $c_i^\dagger$ ($c_i$) are fermionic creation (annihilation) operators at site $i$ (spin indices suppressed for simplifying notation), and $n_{i} = c_{i}^\dagger c_{i}$ is the particle number operator. 
	
	We first introduce the necessary notation. Let $\ket{\mathbf{s}} = \ket{s_1 s_2 \cdots s_N}$ denote a configuration in the occupation number basis, with $s_i = 0$ or $1$ for spinless fermions at site $i$. For an $M$-particle $\ket{\mathbf{s}}$ with an occupied site $i$ ($s_i=1$), define $\bar{\mathbf{s}}$ as the background configuration of the remaining $M-1$ electrons. Each pair $(i,\bar{\mathbf{s}})$ uniquely labels a configuration  $\ket{\mathbf{s}}$. For example, if $\ket{\mathbf{s}} = \ket{110101}$, then for $i = 2$, $\ket{\bar{\mathbf{s}}} = \ket{100101}$.
	
	A general fermionic wavefunction for an $M$-electron system is $|\Phi\rangle = \sum_{\mathbf{s}} W(\mathbf{s}) |\mathbf{s}\rangle$. For a backflow wavefunction expressed as a Slater determinant, the amplitude $W(\mathbf{s})$ can be obained by expanding the determinant along the $i$-th column for any occupied site $i$:
	\begin{equation}
		W(\mathbf{s}) =\det(\Psi) =\sum_{m} \psi_m(i,\bar{\mathbf{s}}) C_m(i,\bar{\mathbf{s}}),
		\label{eq:backflow}
	\end{equation}
	where the matrix element $\Psi_{mi}=\psi_m(i,\bar{\mathbf{s}})$ is the backflow orbital, and $C_m(i,\bar{\mathbf{s}})$ is the corresponding $(m,i)$-cofactor of the matrix $\Psi$~\cite{feynmanEnergySpectrumExcitations1956,sorella2008backflow,tocchioBackflowCorrelationsHubbard2011}. The index $m=1,\dots, M$ labels the orbitas. Unlike the Hartree-Fock approach, where orbitals depend only on single-particle coordinates, here each orbital $\psi_m(i,\bar{\mathbf{s}})$ depends on the full background configuration of the remaining electrons.

    The energy function $E = \langle\Phi|H|\Phi\rangle/\langle\Phi|\Phi\rangle$ is typically optimized variationally with respect to the backflow orbital parameters $\psi_m(i,\bar{\mathbf{s}})$. To gain insight into the structure of these orbitals, we motivate an effective single-particle-like eigenvalue problem (see Supplemental Material):
	\begin{equation}
		\sum_{j} \left[H_m^{\mathrm{eff}}(\bar{\mathbf{s}})\right]_{ij} \psi_m(j,\bar{\mathbf{s}}) = E_m(\bar{\mathbf{s}}) \psi_m(i,\bar{\mathbf{s}}),
		\label{Eq:local_eigenproblem}
	\end{equation}
	where the effective Hamiltonian elements read
	\begin{equation}
		\left[H_m^{\mathrm{eff}}(\bar{\mathbf{s}})\right]_{ij} = \frac{1}{\gamma_m(i,\bar{\mathbf{s}})}\frac{C_m(j,\bar{\mathbf{s}})}{C_m(i,\bar{\mathbf{s}})}\left( t_{ij}\eta^{\mathrm{h}}_{ij}(\bar{\mathbf{s}}) + \delta_{ij}\sum_{k} U_{jk}\eta^{\mathrm{I}}_{jk}(\bar{\mathbf{s}})\right), 
		\label{Eq:H_eff} 
	\end{equation}
	where $E_m(\bar{\mathbf{s}})$ plays the role of an effective energy eigenvalue, $\gamma_m(i,\bar{\mathbf{s}})$ is a nonzero factor, and $\eta_{ij}^{\mathrm{h}}(\bar{\mathbf{s}})$ and $\eta^{\mathrm{I}}_{jk}(\bar{\mathbf{s}})$ enforce that the hopping and interaction terms contribute only when the required sites are occupied appropriately. We restrict to configurations with $C_m(i, \bar{\mathbf{s}}) \neq 0$, as others do not contribute. See Supplemental Material for details.
       
    \begin{figure*}
	\centering
	\begin{minipage}[b]{0.32\textwidth}
		\centering
		\begin{overpic}[width=1\textwidth]{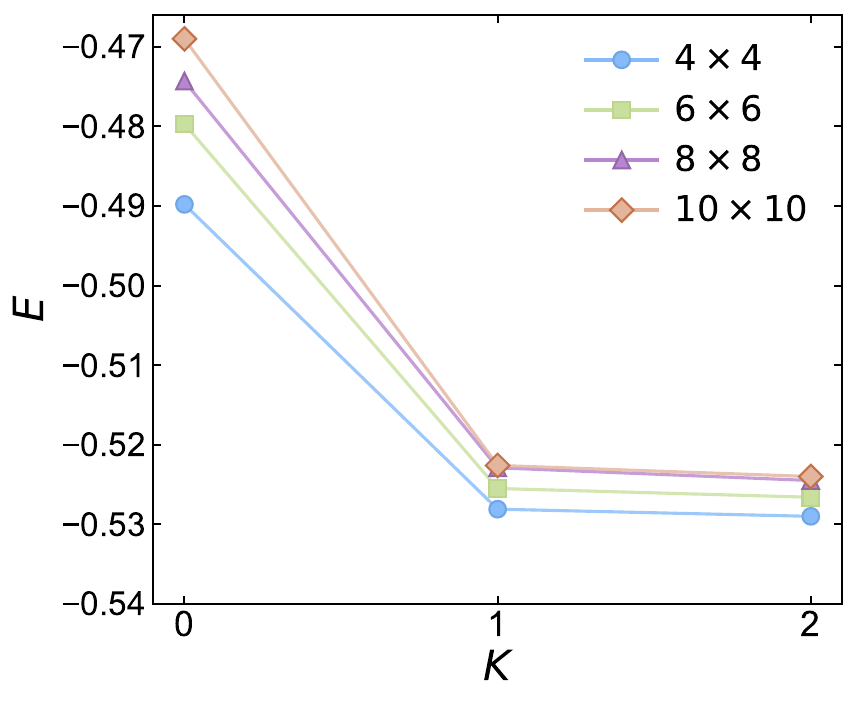}
			\put(5, 85){(a) }   
		\end{overpic}
	\end{minipage}
	\hspace{1pt}
		\begin{minipage}[b]{0.32\textwidth}
		\centering
		\begin{overpic}[width=1\textwidth]{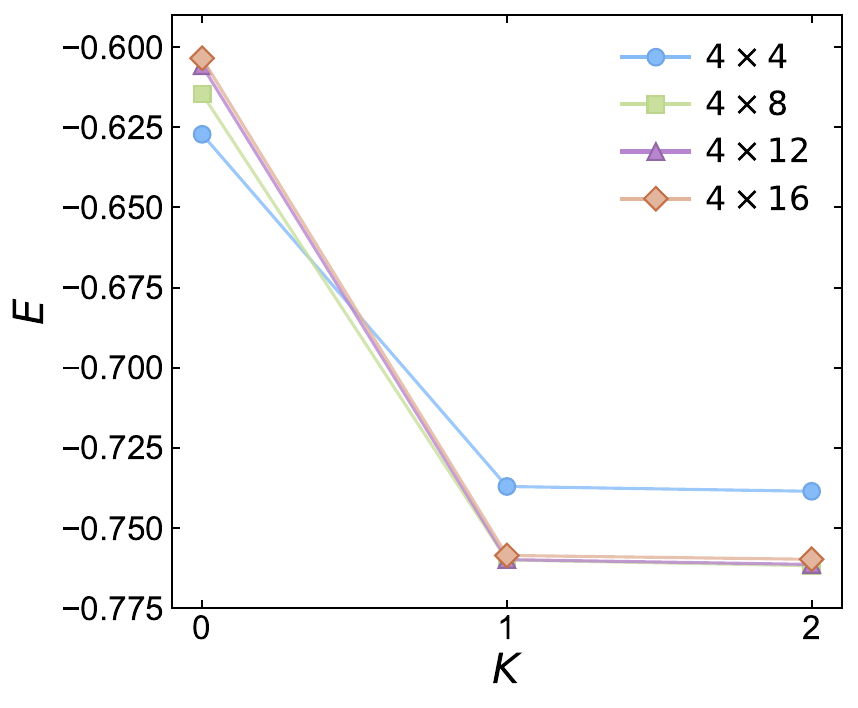}
			\put(5, 85){(b) }  
		\end{overpic}
	\end{minipage}
	\hspace{1pt}
	\begin{minipage}[b]{0.32\textwidth}
		\centering
		\begin{overpic}[width=1\textwidth]{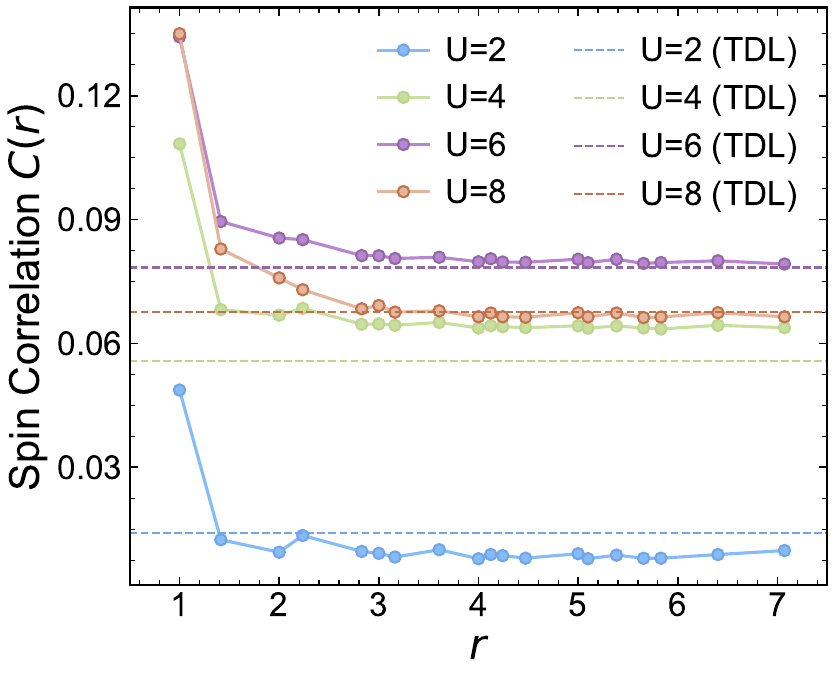}
			\put(5, 85){(c)}  
		\end{overpic}
	\end{minipage}
	\caption{ Energies of the two-dimensional Hubbard model under PBC with $U=8
    $ as a function of the path depth $K$ ($K=0,1,2$) at (a) half-filling and (b) $n_h=0.125$ hole doping. $K=0$ corresponds to the Hartree-Fock state. Increasing $K$ systematically lowers the energy. (c) Comparison of staggered spin correlation functions $C(r)=(-1)^{x+y} \langle \mathbf{S}(0,0) \cdot \mathbf{S}(x,y) \rangle $ as a function of distance $r = \sqrt{x^2+y^2}$ for $10\times 10$ Hubbard model at half-filling under PBC for HB $K=1$. The dashed horizontal line denotes the thermodynamic limit (TDL) value evaluated from  AFQMC~\cite{qinBenchmarkStudyTwodimensional2016}. }
    \label{fig:HB_family_energy_correlation}
\end{figure*}
    
Note Eq.~(\ref{Eq:H_eff}) is nonlinear and computationally challenging because $H^{\mathrm{eff}}_m$ depends on the configuration of all other sites. To elucidate its structure, we make a local kernel approximation that retains only two-site contributions in which $[H^{\mathrm{eff}}_m(\bar{\mathbf{s}})]_{ij}$ depends only on the occupation variable at site $i$ and its nearest neighbor $j$. Under this approximation, the effective Hamiltonian reduces to a local kernel $H^{\mathrm{loc}}_m(i,s_i;j,s_j)$ determined solely by the two site occupations.

This kernel defines local transitions between sites, with each transition element depending on the local variables $s_i$ and $s_j$. Iterating such kernel actions on an initial orbital (e.g., the Hartree-Fock orbital) therefore accumulates background information step by step, generating a nonlocal dependence on the full configuration, a natural manifestation of backflow correlations. The iterative structure can be expressed as a sum over paths of successive local transitions, reminiscent of the power method. 

Motivated by this observation, we introduce a family of variational fermionic states termed hierarchical backflow (HB) wavefunctions, by parameterizing the backflow orbitals as a path expansion with depth $K$:
\begin{equation}
    \psi_{m}^{\mathrm{HB}}(i, \bar{\mathbf{s}}) = \sum_{i_1, \dots, i_K} \; \prod_{l=1}^{K} f_m^{[l]}(i_{l-1}, s_{i_{l-1}}; i_{l}, s_{i_l}),
    \label{eq:psi_loc}
\end{equation}
with the initial site $i \equiv i_0$. Each factor $f_m^{[l]}$ contains a set of variational parameters encoding how the orbital at the reference site $i_{l-1}$ is influenced by the occupations of its neighboring sites $i_l$, thereby generating backflow correlations. The path depth $K$ defines the range of backflow correlations from a starting site $i$ and increasing $K$ systematically extends this range, providing a variational ansatz with controlled and systematically improvable expressive power.

For $K=0$, the expansion reduces to a function depending only on $i$ (i.e., $\psi_m^{\mathrm{HB}}(i,\bar{\mathbf{s}})\big|_{K=0} = \psi^{\mathrm{HF}}_m(i)$). This recovers the Hartree-Fock wavefunction, where the orbital is independent of the background $\bar{\mathbf{s}}$ and thus lack backflow correlations. The first-order expansion ($K=1$) reads
	\begin{equation}
		\psi_{m}^{\mathrm{HB}}(i,\bar{\mathbf{s}})\big|_{K=1}=\sum_{j} f_m^{[1]}(i,s_i; j,s_j),
	\end{equation}
	and the second-order expansion ($K=2$)  is
	\begin{equation}
		\psi_{m}^{\mathrm{HB}}(i,\bar{\mathbf{s}})\big|_{K=2}=  \sum_{j,p} f_m^{[1]}(i,s_i; j,s_{j})f_m^{[2]}(j,s_{j}; p,s_{p}),
	\end{equation}
where $j$ runs over $i$ and its four nearest neighbors, and $p$ runs over $j$ and its four nearest neighbors.

   The wavefunction amplitude is then obtained as a Slater determinant of the HB orbitals, $W(\mathbf{s}) = \det\left[ \psi_m^{\mathrm{HB}}(i,\bar{\mathbf{s}}) \right]$. Notably, despite the small path depth $K$, the Slater determinant couples all HB orbitals, generating many-body correlations in the amplitude $W(\mathbf{s})$. 

  One can also relax the two‑site kernel approximation to higher‑order local kernels, e.g., three-site kernels, leading to a three‑site variant of the hierarchical backflow ansatz with factors $f_m^{[l]}(i,s_i;\; j,s_j;\; p,s_p)$. We expect that the two‑site HB family with large $K$  have similar expressivity to the three‑site family with small $K$; for instance, the two-site construction with $K=2$ closely resembles the three-site construction with $K=1$ (up to a reparameterization). Here we always adopt the two‑site HB family.

\textit{Effectiveness of the hierarchical backflow.} We first examine the convergence of the HB ansatz with respect to the path depth $K$.  All calculations are performed using variational Monte Carlo with gradient‑based optimization such as stochastic reconfiguration~\cite{SR1998}. Figure~\ref{fig:HB_family_energy_correlation}(a) shows the $K$‑dependence of the energy for the half‑filled Hubbard model on system sizes from $4\times4$ to $10\times10$ with $U=8$ under periodic boundary conditions (PBC), while Figure~\ref{fig:HB_family_energy_correlation}(b) shows the analogous results at hole doping $n_h=0.125$ for $4\times L$ systems ($L=4$ to $16$). We show results for $K=0, 1, 2$. In all cases, the energy drops sharply from $K=0$ (Hartree‑Fock) to $K=1$, and then improves modestly from $K=1$ to $K=2$. Remarkably, for half‑filling, even $K=1$ achieves relative energy errors as low as around $0.5\%$ across all system sizes from $4\times4$ to $10\times10$ and interaction strengths $U=2-8$, when benchmarked against unbiased auxiliary‑field quantum Monte Carlo (AFQMC) reference energies (see Table~\ref{table:hubbard_errors}). This rapid convergence shows that the locality‑induced hierarchy captures the vast majority of many‑body correlations already at the shallow nontrivial depth. For the doped case, our $K=1$ results already outperform existing neural‑network‑based backflow ansatze~\cite{luo2019backflow,fermionNN2022} (see later), and increasing $K$ systematically pushes the energy even lower, which is a direct manifestation of the systematic improvability promised by our framework.

To further assess the quality of the correlations captured by the HB, we compute the staggered spin‑spin correlation function $C(r)=(-1)^{x+y}\langle \mathbf{S}(0,0)\cdot\mathbf{S}(x,y) \rangle$, where $r=\sqrt{x^2+y^2}$ is the distance and $\mathbf{S}(x,y)$ is the spin operator at coordinate $(x,y)$. Figure~\ref{fig:HB_family_energy_correlation}(c) shows its behavior on a $10\times10$ lattice. As a reference, we use the thermodynamic limit (TDL) magnetic moments from AFQMC evaluation~\cite{qinBenchmarkStudyTwodimensional2016}. Our $10\times10$ results exhibit a similar $U$‑dependence to the TDL values. For instance, correlations are strongest at $U=6$, and at both $U=6$ and $U=8$ they display clear long‑range order, consistent with the well‑known antiferromagnetic ground state at half‑filling. Strikingly, all of this qualitatively correct physics, including the correct interaction strength for maximal order, is already captured by the $K=1$ wavefunction.

	\begin{table}[htbp]
		\centering
		\caption{Relative energy errors for the Hubbard model at half-filling under PBC with hierarchical backflow $K=1$ for various $U$ and system sizes, and all errors are below 0.65\%. Energy values are listed in Supplemental Material. AFQMC energies are used for references~\cite{qinBenchmarkStudyTwodimensional2016}.}
		\label{table:hubbard_errors}
        \begin{tabular*}{\hsize}{@{}@{\extracolsep{\fill}}ccccc@{}}
			\hline\hline
			size & \textbf{$U=2$} & \textbf{$U=4$} & \textbf{$U=6$} & \textbf{$U=8$} \\
			\hline
			$4\times4$   & 0.0015 & 0.0029 & 0.0017 & 0.0032 \\
			$6\times6$   & 0.0009 & 0.0023 & 0.0042 & 0.0044 \\
			$8\times8$   & 0.0009 & 0.0024 & 0.0051 & 0.0064 \\
			$10\times10$ & 0.0012 & 0.0031 & 0.0046 & 0.0053 \\
			\hline\hline
		\end{tabular*}
	\end{table}

	\begin{figure}[t]
		\centering
		\includegraphics[width=1\columnwidth]{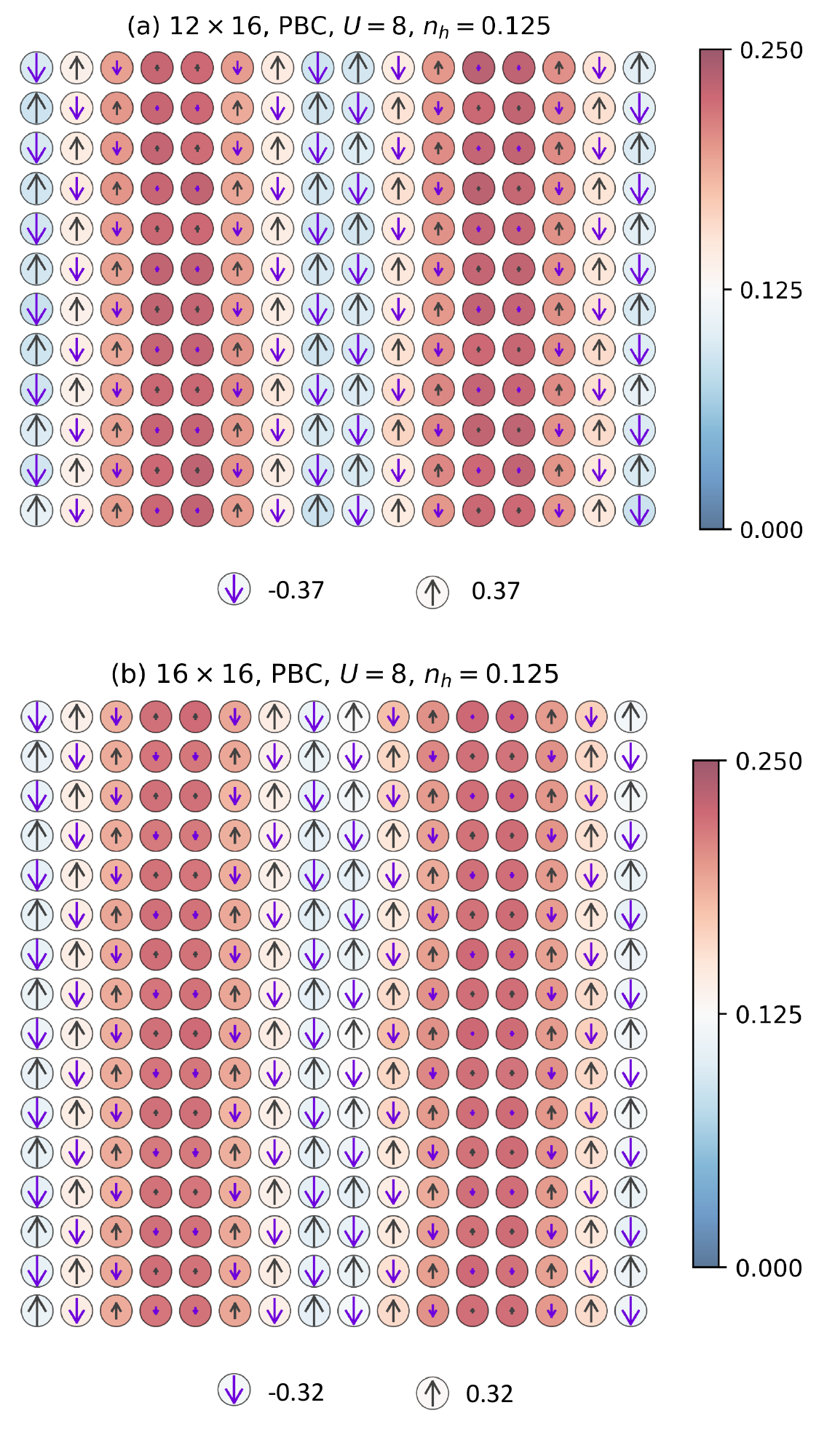}
		\caption{The distribution of hole density (color map) and local spin $S_z$ moment (arrow length proportional to magnitude) on (a) $12\times 16$ and (b) $16\times 16$ PBC lattice at  $n_h=0.125$ hole doping with $U=8$.}
		\label{fig:stripe}
	\end{figure}

We then scale to larger systems, including $12\times16$ and $16\times16$ lattices at hole doping $n_h=0.125$ and $U=8$ under PBC. The ground state energies on the $16\times16$ lattice are $-0.5889$ (HB $K=0$), $-0.7510$ (HB $K=1$) and $-0.7526$ (HB $K=2$), only about $0.2\%$ higher than the recent variational auxiliary‑field Monte Carlo result of Sorella {\it et al.}, $-0.7544$~\cite{sorella2023,doi:10.1126/science.adg9774}. Figure~\ref{fig:stripe} presents the hole density and spin moment distributions using $K=2$, which clearly reveal a stripe phase with wavelength $\lambda = 8$, consistent with previous studies~\cite{doi:10.1126/science.aam7127,liu2025hubbard}. These results demonstrate that HB scales efficiently to large system sizes ($12\times16$, $16\times16$), systematically improves the energy with $K$, and captures the stripe phase with wavelength $\lambda=8$, consistent with prior studies~\cite{zheng2017stripe,liu2025hubbard}. 

Overall, these results establish HB as a systematically improvable and highly efficient approach for large‑scale simulations of strongly correlated fermions.

\textit{Residual Hierarchical Backflow.} The HB framework naturally enables a local-nonlocal decomposition, providing a direct bridge to neural network quantum states while preserving interpretable representation. By keeping the path depth $K$ small and incorporating residual correlations via an additional nonlocal factor, we construct a variant termed residual hierarchical backflow (RHB). The nonlocal part can be parameterized by a flexible function approximator (e.g., a neural network or a tensor network), while the HB backbone remains physically transparent and systematically improvable.

A convenient realization of RHB uses a multi‑determinant construction in which all Slater determinants share the same HB orbitals $\psi_m^{\mathrm{HB}}$, but each determinant carries its own nonlocal factor $Q_m^{[\alpha]}(\bar{\mathbf{s}})$:
\begin{equation}
	\Psi^{[\alpha]}_{mi} = \psi_m^{\mathrm{HB}}(i, \bar{\mathbf{s}}) \, Q_m^{[\alpha]}(\bar{\mathbf{s}}), 
	\label{Eq:finalform}
\end{equation}
and thus the amplitude is $W(\mathbf{s})=\sum_{\alpha} \det(\Psi^{[\alpha]})$, where $\alpha$ labels the Slater determinant.

Compared to conventional backflow architectures that rely on a single global network to directly generate orbital functions~\cite{luo2019backflow,fermionNN2022,clark2024unifying}, the RHB representation is more compact (see End Matter) and can be optimized efficiently via a two‑stage scheme (details in Supplemental Material). As an example, we instantiate the RHB with an HB $K=2$ backbone and a feed‑forward neural network (FNN) parameterizing $Q_m^{[\alpha]}(\bar{\mathbf{s}})$. On the $16\times16$ lattice at $n_h=0.125$ and $U=8$, RHB with the FNN yields a ground state energy of $-0.7543(2)$, in excellent agreement with the aforementioned variational auxiliary‑field Monte Carlo result $-0.7544$~\cite{sorella2023,doi:10.1126/science.adg9774}, while the spin moment and hole density distributions remain essentially unchanged from the pure HB $K=2$ results.

\begin{figure}
		\centering
		\includegraphics[width=1\columnwidth]{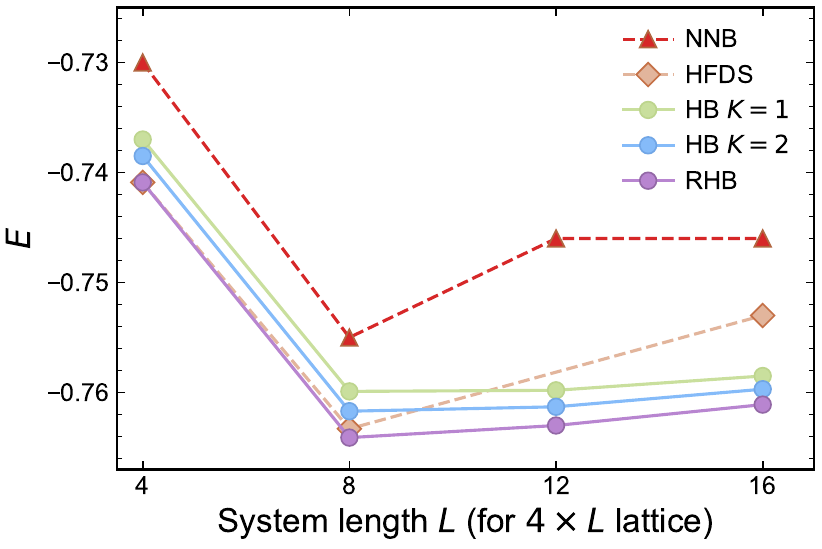}
		\caption{Energy comparison with NNB~\cite{luo2019backflow} and HFDS~\cite{fermionNN2022} for different sizes at $n_h=0.125$ and $U=8$, including HB $K=1$, $K=2$ and the RHB.}
		\label{fig:4xlcompare}
	\end{figure}

We further compare our approach with other backflow‑based methods that employ a single global network: neural‑network backflow (NNB)~\cite{luo2019backflow} and hidden‑fermion determinant state (HFDS)~\cite{fermionNN2022}. All calculations are performed on $4\times L$ lattices at $n_h=0.125$ and $U=8$, shown in Figure~\ref{fig:4xlcompare}. For the $4\times4$ system, both our RHB and HFDS agree excellently with exact diagonalization, with energy errors of approximately $0.0008$. On $4\times8$, our RHB energy is slightly lower than that of HFDS, and the improvement becomes more pronounced on larger lattices (e.g., $4\times16$). Remarkably, even the simplest HB with $K=1$ already outperforms NNB across all system sizes, and increasing to $K=2$ or using RHB yields further improvements. These results highlight the flexibility of the HB framework and the advantage of its local‑nonlocal structure.

\textit{Conclusions.} Although the backflow wavefunction was proposed about 70 years ago, its underlying organizing principle has remained elusive. In this work, we reveal that locality provides such a natural organizing principle, inducing a hierarchical structure controlled by a path depth $K$. The resulting HB wavefunction smoothly interpolates from the Hartree-Fock state ($K=0$) to increasingly correlated states ($K>0$), offering a clear physical interpretation of backflow correlations. It also admits a local-nonlocal decomposition, where the nonlocal component can be chosen flexibly, such as a convolutional neural network, a tensor network or a Jastrow factor. This flexibility makes HB a versatile backbone for designing expressive yet interpretable variational states.

The HB scheme applies directly to many two dimensional fermionic models, as well as to three dimensional lattices and quantum chemistry problems. Crucially, its expressive power is systematically improvable by simply increasing the path depth $K$, offering a new framework with interpretability for large‑scale simulations of strongly correlated fermions.

	\textit{Acknowledgment.}---Y.-T. Zhou and Z.-W. Zhou are supported by National Natural Science Foundation of China (Grant No.12474366) and Innovation Program for Quantum Science and Technology (Grant No.2021ZD0301900). W.-Y. Liu is supported by National Natural Science Foundation of China (Grant No.12534009). The computational resources utilized in this work were supported by SCNet Supercomputing Network.

	\section{End matter}

    \textit{Variational parameters in the HB.} The HB wavefunction is built from elementary variational tensors $f_m^{[l]}(i,s_i;j,s_j)$ according to Eq.~\eqref{eq:psi_loc}. Here we take $K=2$ as an illustration,
\begin{equation}
	\psi_{m}^{\mathrm{HB}}(i,\bar{\mathbf{s}})\big|_{K=2}= \sum_{j,p} f_m^{[1]}(i,s_i;j,s_{j}) \, f_m^{[2]}(j,s_{j};p,s_{p}),
\end{equation}
where $j$ runs over sites nearby $i$ (typically including the site itself and its four nearest neighbors on a square lattice, giving $z=5$ possible $j$ values), and $p$ runs over sites nearby $j$.  All spin indices in the $f$ tensors (including $s_i$, $s_p$ and $s_j$) are directly specified by the configuration $\ket{{\bf s}}$, taking one of four possibilities for spinful fermions: $\ket{0}$, $\ket{\uparrow}$, $\ket{\downarrow}$, or $\ket{\uparrow\downarrow}$. 

In practice, the sums are evaluated on the fly according to $\ket{\mathbf{s}}$, and the resulting matrix $\Psi_{mi}^{\mathrm{HB}} = \psi_m^{\mathrm{HB}}(i,\bar{\mathbf{s}})$ is used to compute $W(\mathbf{s}) = \det(\Psi^{\mathrm{HB}})$. The total number of variational parameters scales as $zd^2MNK$, where $d=4$ is the local Hilbert space dimension, $M$ the number of electrons, and $N$ the total number of lattice sites.

	\begin{figure}
		\centering
		\includegraphics[width=1\columnwidth]{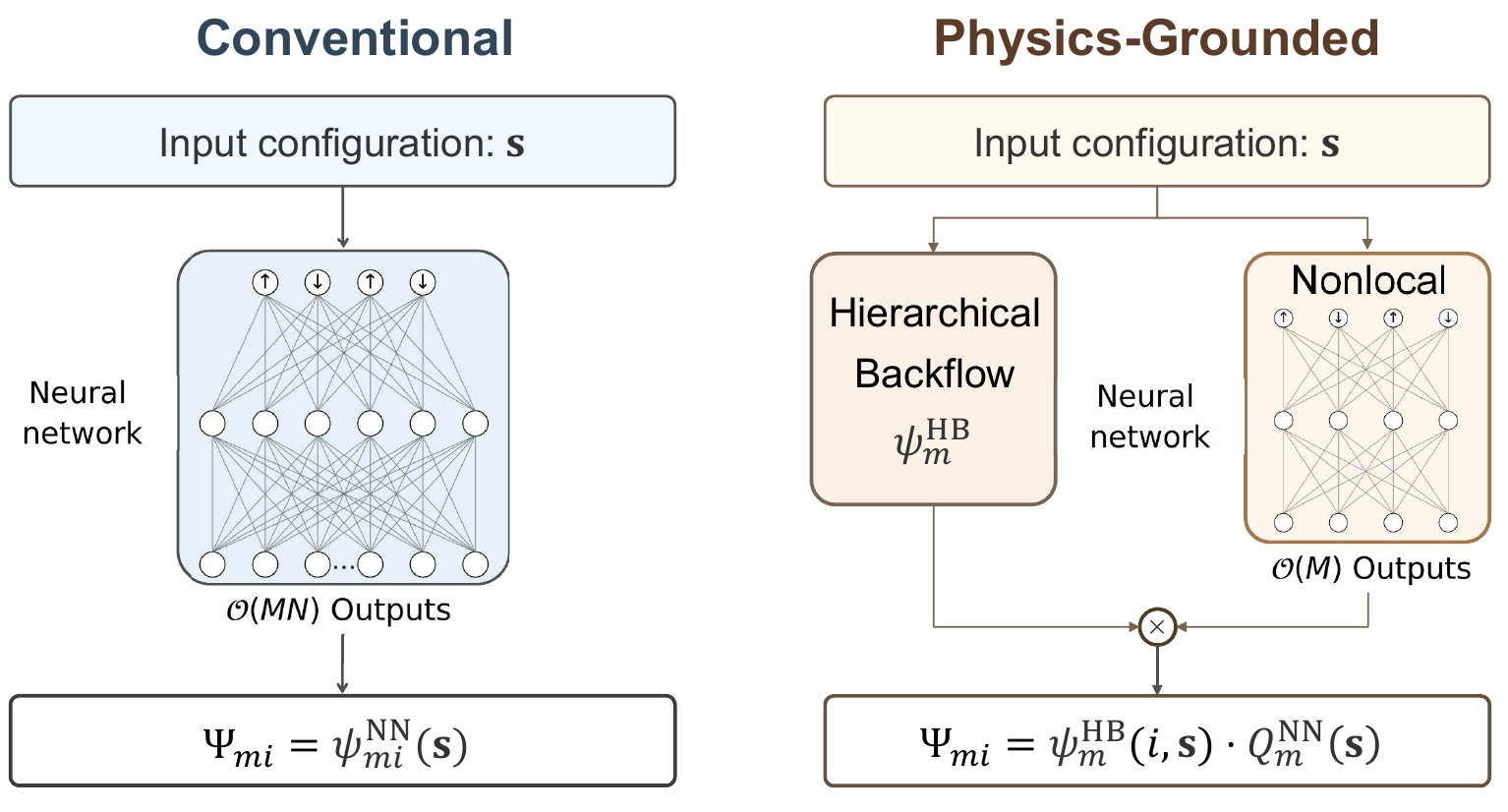}
		\caption{ Conventional neural network states employ a global architecture (left), whereas our physics-grounded ansatz features a local-nonlocal structure (right). For the same number of determinants, which provides similar expressive power, the number of variational parameters in our representation scales as $O(N_{\text{neuron}}N) + O(N^2)$, favorable compared to $O(N_{\text{neuron}}N^2)$ for global architectures.}  
		\label{fig:nnbf-vs-decomposed-approach-comparison}
	\end{figure}
	
	 \textit{RHB as a compact representation.} In the RHB, the neural network only needs to output $M$ values per determinant to form the Slater determinant in combination with the HB(see Figure~\ref{fig:nnbf-vs-decomposed-approach-comparison}). Thus, the parameter count scales as  $O(N_{\text{neuron}} M) + O(MN)$, where the first term comes from the neural network and the second from the HB backbone. Here $N_{\text{neuron}}$ is the number of hidden neurons that directly connects to the $M$ outputs (each connection contributes parameters).

    In contrast, conventional global network architectures require $O(MN)$ output values to construct each Slater determinant~\cite{luo2019backflow,fermionNN2022,clark2024unifying}, leading to a parameter scaling of $O(N_{\text{neuron}} M N)$. Typically $M \propto N$ (e.g., $M = \frac{7}{8}N$ for hole doping $n_h=0.125$). 
    Hence, for the same number of determinants which provides similar expressive power, the RHB ansatz has a favorable scaling: $O(N_{\text{neuron}} N) + O(N^2)$ versus $O(N_{\text{neuron}} N^2)$ for global architectures.
	
	\bibliography{ref} 

@article{Liang2026,
  title = {Investigating the Fermi-Hubbard model by the tensor-backflow method},
  author = {Liang, Xiao},
  journal = {Phys. Rev. B},
  volume = {113},
  issue = {15},
  pages = {155121},
  numpages = {13},
  year = {2026},
  month = {Apr},
  publisher = {American Physical Society},
  doi = {10.1103/dqc7-zsvc},
  url = {https://link.aps.org/doi/10.1103/dqc7-zsvc}
}

@article{Jastrow1955,
  title = {Many-Body Problem with Strong Forces},
  author = {Jastrow, Robert},
  journal = {Phys. Rev.},
  volume = {98},
  issue = {5},
  pages = {1479--1484},
  numpages = {0},
  year = {1955},
  month = {Jun},
  publisher = {American Physical Society},
  doi = {10.1103/PhysRev.98.1479},
  url = {https://link.aps.org/doi/10.1103/PhysRev.98.1479}
}

@article{Jain1992jastrow,
  title = {Jastrow-Slater trial wave functions for the fractional quantum Hall effect: Results for few-particle systems},
  author = {Dev, Gautam and Jain, J. K.},
  journal = {Phys. Rev. B},
  volume = {45},
  issue = {3},
  pages = {1223--1230},
  numpages = {0},
  year = {1992},
  month = {Jan},
  publisher = {American Physical Society},
  doi = {10.1103/PhysRevB.45.1223},
  url = {https://link.aps.org/doi/10.1103/PhysRevB.45.1223}
}

@article{verstraete2008,
    author = {Verstraete, F. and Murg, V. and Cirac, J. I.},
    title = {Matrix product states, projected entangled pair states, and variational renormalization group methods for quantum spin systems},
    journal = {Advances in Physics},
    volume = {57},
    number = {2},
    pages = {143-224},
    year = {2008},
    doi = {10.1080/14789940801912366},
}

@article{sorella2023,
  title = {Systematically improvable mean-field variational ansatz for strongly correlated systems: Application to the Hubbard model},
  author = {Sorella, Sandro},
  journal = {Phys. Rev. B},
  volume = {107},
  issue = {11},
  pages = {115133},
  numpages = {17},
  year = {2023},
  month = {Mar},
  publisher = {American Physical Society},
  doi = {10.1103/PhysRevB.107.115133},
  url = {https://link.aps.org/doi/10.1103/PhysRevB.107.115133}
}

@article{
doi:10.1126/science.adg9774,
author = {Dian Wu  et al.  },
title = {Variational benchmarks for quantum many-body problems},
journal = {Science},
volume = {386},
number = {6719},
pages = {296-301},
year = {2024},
doi = {10.1126/science.adg9774},
}

@article{PhysRevX.14.021030,
  title = {Neural Wave Functions for Superfluids},
  author = {Lou, Wan Tong and Sutterud, Halvard and Cassella, Gino and Foulkes, W. M. C. and Knolle, Johannes and Pfau, David and Spencer, James S.},
  journal = {Phys. Rev. X},
  volume = {14},
  issue = {2},
  pages = {021030},
  numpages = {19},
  year = {2024},
  month = {May},
  publisher = {American Physical Society},
  doi = {10.1103/PhysRevX.14.021030},
  url = {https://link.aps.org/doi/10.1103/PhysRevX.14.021030}
}

@article{SR1998,
  title = {Green Function Monte Carlo with Stochastic Reconfiguration},
  author = {Sorella, Sandro},
  journal = {Phys. Rev. Lett.},
  volume = {80},
  issue = {20},
  pages = {4558--4561},
  numpages = {0},
  year = {1998},
  month = {May},
  publisher = {American Physical Society},
  doi = {10.1103/PhysRevLett.80.4558},
  url = {https://link.aps.org/doi/10.1103/PhysRevLett.80.4558}
}

@article{luo2019backflow,
  title = {Backflow Transformations via Neural Networks for Quantum Many-Body Wave Functions},
  author = {Luo, Di and Clark, Bryan K.},
  journal = {Phys. Rev. Lett.},
  volume = {122},
  issue = {22},
  pages = {226401},
  numpages = {6},
  year = {2019},
  month = {Jun},
  publisher = {American Physical Society},
  doi = {10.1103/PhysRevLett.122.226401},
  url = {https://link.aps.org/doi/10.1103/PhysRevLett.122.226401}
}

@article{clark2024unifying,
  title = {Unifying view of fermionic neural network quantum states: From neural network backflow to hidden fermion determinant states},
  author = {Liu, Zejun and Clark, Bryan K.},
  journal = {Phys. Rev. B},
  volume = {110},
  issue = {11},
  pages = {115124},
  numpages = {17},
  year = {2024},
  month = {Sep},
  publisher = {American Physical Society},
  doi = {10.1103/PhysRevB.110.115124},
  url = {https://link.aps.org/doi/10.1103/PhysRevB.110.115124}
}

@article{fermionNN2022,
author = {Javier Robledo Moreno  and Giuseppe Carleo  and Antoine Georges  and James Stokes },
title = {Fermionic wave functions from neural-network constrained hidden states},
journal = {Proceedings of the National Academy of Sciences},
volume = {119},
number = {32},
pages = {e2122059119},
year = {2022},
doi = {10.1073/pnas.2122059119},
URL = {https://www.pnas.org/doi/abs/10.1073/pnas.2122059119}
}

@article{DMFT1,
  title = {Dynamical mean-field theory of strongly correlated fermion systems and the limit of infinite dimensions},
  author = {Georges, Antoine and Kotliar, Gabriel and Krauth, Werner and Rozenberg, Marcelo J.},
  journal = {Rev. Mod. Phys.},
  volume = {68},
  issue = {1},
  pages = {13--125},
  numpages = {0},
  year = {1996},
  month = {Jan},
  publisher = {American Physical Society},
  doi = {10.1103/RevModPhys.68.13},
  url = {https://link.aps.org/doi/10.1103/RevModPhys.68.13}
}

@article{DMFT2,
  title = {Electronic structure calculations with dynamical mean-field theory},
  author = {Kotliar, G. and Savrasov, S. Y. and Haule, K. and Oudovenko, V. S. and Parcollet, O. and Marianetti, C. A.},
  journal = {Rev. Mod. Phys.},
  volume = {78},
  issue = {3},
  pages = {865--951},
  numpages = {0},
  year = {2006},
  month = {Aug},
  publisher = {American Physical Society},
  doi = {10.1103/RevModPhys.78.865},
  url = {https://link.aps.org/doi/10.1103/RevModPhys.78.865}
}

@article{cDMFT,
  title = {Cellular Dynamical Mean Field Approach to Strongly Correlated Systems},
  author = {Kotliar, Gabriel and Savrasov, Sergej Y. and P\'alsson, Gunnar and Biroli, Giulio},
  journal = {Phys. Rev. Lett.},
  volume = {87},
  issue = {18},
  pages = {186401},
  numpages = {4},
  year = {2001},
  month = {Oct},
  publisher = {American Physical Society},
  doi = {10.1103/PhysRevLett.87.186401},
  url = {https://link.aps.org/doi/10.1103/PhysRevLett.87.186401}
}

@article{CPAFQMC1,
  title = {Constrained path Monte Carlo method for fermion ground states},
  author = {Zhang, Shiwei and Carlson, J. and Gubernatis, J. E.},
  journal = {Phys. Rev. B},
  volume = {55},
  issue = {12},
  pages = {7464--7477},
  numpages = {0},
  year = {1997},
  month = {Mar},
  publisher = {American Physical Society},
  doi = {10.1103/PhysRevB.55.7464},
  url = {https://link.aps.org/doi/10.1103/PhysRevB.55.7464}
}

@article{CPAFQMC2,
  title = {Quantum Monte Carlo Method using Phase-Free Random Walks with Slater Determinants},
  author = {Zhang, Shiwei and Krakauer, Henry},
  journal = {Phys. Rev. Lett.},
  volume = {90},
  issue = {13},
  pages = {136401},
  numpages = {4},
  year = {2003},
  month = {Apr},
  publisher = {American Physical Society},
  doi = {10.1103/PhysRevLett.90.136401},
  url = {https://link.aps.org/doi/10.1103/PhysRevLett.90.136401}
}

@article{DiaMC,
title = {Diagrammatic Monte Carlo},
journal = {Physics Procedia},
volume = {6},
pages = {95-105},
year = {2010},
note = {Computer Simulations Studies in Condensed Matter Physics XXI},
issn = {1875-3892},
doi = {https://doi.org/10.1016/j.phpro.2010.09.034},
url = {https://www.sciencedirect.com/science/article/pii/S1875389210006498},
author = {Kris {Van Houcke} and Evgeny Kozik and N. Prokof’ev and B. Svistunov}
}

@article{DDMC,
  title = {Determinant Diagrammatic Monte Carlo Algorithm in the Thermodynamic Limit},
  author = {Rossi, Riccardo},
  journal = {Phys. Rev. Lett.},
  volume = {119},
  issue = {4},
  pages = {045701},
  numpages = {5},
  year = {2017},
  month = {Jul},
  publisher = {American Physical Society},
  doi = {10.1103/PhysRevLett.119.045701},
  url = {https://link.aps.org/doi/10.1103/PhysRevLett.119.045701}
}

@article{GFQMC,
  title = {Currents and Green's functions of impurities out of equilibrium: Results from inchworm quantum Monte Carlo},
  author = {Antipov, Andrey E. and Dong, Qiaoyuan and Kleinhenz, Joseph and Cohen, Guy and Gull, Emanuel},
  journal = {Phys. Rev. B},
  volume = {95},
  issue = {8},
  pages = {085144},
  numpages = {10},
  year = {2017},
  month = {Feb},
  publisher = {American Physical Society},
  doi = {10.1103/PhysRevB.95.085144},
  url = {https://link.aps.org/doi/10.1103/PhysRevB.95.085144}
}

@article{jiang2019,
author = {Hong-Chen Jiang  and Thomas P. Devereaux },
title = {Superconductivity in the doped \text{Hubbard} model and its interplay with next-nearest hopping $t'$},
journal = {Science},
volume = {365},
number = {6460},
pages = {1424-1428},
year = {2019},
doi = {10.1126/science.aal5304},
URL = {https://www.science.org/doi/abs/10.1126/science.aal5304}
}

@article{gong2021robust,
  title = {Robust $d$-Wave Superconductivity in the Square-Lattice $t\text{\ensuremath{-}}J$ Model},
  author = {Gong, Shoushu and Zhu, W. and Sheng, D. N.},
  journal = {Phys. Rev. Lett.},
  volume = {127},
  issue = {9},
  pages = {097003},
  numpages = {6},
  year = {2021},
  month = {Aug},
  publisher = {American Physical Society},
  doi = {10.1103/PhysRevLett.127.097003},
  url = {https://link.aps.org/doi/10.1103/PhysRevLett.127.097003}
}

@article{li2023tangent,
  title = {Tangent Space Approach for Thermal Tensor Network Simulations of the 2D Hubbard Model},
  author = {Li, Qiaoyi and Gao, Yuan and He, Yuan-Yao and Qi, Yang and Chen, Bin-Bin and Li, Wei},
  journal = {Phys. Rev. Lett.},
  volume = {130},
  issue = {22},
  pages = {226502},
  numpages = {8},
  year = {2023},
  month = {Jun},
  publisher = {American Physical Society},
  doi = {10.1103/PhysRevLett.130.226502},
  url = {https://link.aps.org/doi/10.1103/PhysRevLett.130.226502}
}

@article{qu2024phase,
  title = {Phase Diagram, $d$-Wave Superconductivity, and Pseudogap of the $t\ensuremath{-}{t}^{\ensuremath{'}}\ensuremath{-}J$ Model at Finite Temperature},
  author = {Qu, Dai-Wei and Li, Qiaoyi and Gong, Shou-Shu and Qi, Yang and Li, Wei and Su, Gang},
  journal = {Phys. Rev. Lett.},
  volume = {133},
  issue = {25},
  pages = {256003},
  numpages = {8},
  year = {2024},
  month = {Dec},
  publisher = {American Physical Society},
  doi = {10.1103/PhysRevLett.133.256003},
  url = {https://link.aps.org/doi/10.1103/PhysRevLett.133.256003}
}

@article{dmrg1992,
  title = {Density matrix formulation for quantum renormalization groups},
  author = {White, Steven R.},
  journal = {Phys. Rev. Lett.},
  volume = {69},
  issue = {19},
  pages = {2863--2866},
  numpages = {0},
  year = {1992},
  month = {Nov},
  publisher = {American Physical Society},
  doi = {10.1103/PhysRevLett.69.2863},
  url = {https://link.aps.org/doi/10.1103/PhysRevLett.69.2863}
}

@article{PhysRevMaterials.3.054605,
  title = {Exciton Mott transition revisited},
  author = {Guerci, Daniele and Capone, Massimo and Fabrizio, Michele},
  journal = {Phys. Rev. Mater.},
  volume = {3},
  issue = {5},
  pages = {054605},
  numpages = {9},
  year = {2019},
  month = {May},
  publisher = {American Physical Society},
  doi = {10.1103/PhysRevMaterials.3.054605},
  url = {https://link.aps.org/doi/10.1103/PhysRevMaterials.3.054605}
}

@article{PhysRevB.70.035114,
  title = {Slave-rotor mean-field theories of strongly correlated systems and the Mott transition in finite dimensions},
  author = {Florens, Serge and Georges, Antoine},
  journal = {Phys. Rev. B},
  volume = {70},
  issue = {3},
  pages = {035114},
  numpages = {15},
  year = {2004},
  month = {Jul},
  publisher = {American Physical Society},
  doi = {10.1103/PhysRevB.70.035114},
  url = {https://link.aps.org/doi/10.1103/PhysRevB.70.035114}
}

@article{doi:10.1142/S0217979292000414,
author = {Fr\'{e}sard, R. and W\"{o}lfle, P.},
title = {Unified Slave Boson Representation of Spin and Charge Degrees of Freedom for Strongly Correlated Fermi Systems},
journal = {International Journal of Modern Physics B},
volume = {06},
number = {05n06},
pages = {685-704},
year = {1992},
doi = {10.1142/S0217979292000414}
}

@article{doi:10.1098/rspa.1963.0204,
author = {Hubbard, J. },
title = {Electron correlations in narrow energy bands},
journal = {Proceedings of the Royal Society of London. Series A. Mathematical and Physical Sciences},
volume = {276},
number = {1365},
pages = {238-257},
year = {1963},
doi = {10.1098/rspa.1963.0204}
}

@article{
doi:10.1126/science.aam7127,
author = {Bo-Xiao Zheng et al.  },
title = {Stripe order in the underdoped region of the two-dimensional Hubbard model},
journal = {Science},
volume = {358},
number = {6367},
pages = {1155-1160},
year = {2017},
doi = {10.1126/science.aam7127}}

@article{annurev:/content/journals/10.1146/annurev-conmatphys-031620-102024,
   author = "Arovas, Daniel P. and Berg, Erez and Kivelson, Steven A. and Raghu, Srinivas",
   title = "The Hubbard Model", 
   journal= "Annu. Rev. Condens. Matter Phys.",
   year = "2022",
   volume = "13",
   number = "Volume 13, 2022",
   pages = "239-274",
   doi = "https://doi.org/10.1146/annurev-conmatphys-031620-102024",
   url = "https://www.annualreviews.org/content/journals/10.1146/annurev-conmatphys-031620-102024",
   publisher = "Annual Reviews",
   issn = "1947-5462",
   type = "Journal Article",
  }

@article{RevModPhys.66.763,
  title = {Correlated electrons in high-temperature superconductors},
  author = {Dagotto, Elbio},
  journal = {Rev. Mod. Phys.},
  volume = {66},
  issue = {3},
  pages = {763--840},
  numpages = {0},
  year = {1994},
  month = {Jul},
  publisher = {American Physical Society},
  doi = {10.1103/RevModPhys.66.763},
  url = {https://link.aps.org/doi/10.1103/RevModPhys.66.763}
}

@article{corboz2010simulation,
  title = {Simulation of strongly correlated fermions in two spatial dimensions with fermionic projected entangled-pair states},
  author = {Corboz, Philippe and Or\'us, Rom\'an and Bauer, Bela and Vidal, Guifr\'e},
  journal = {Phys. Rev. B},
  volume = {81},
  issue = {16},
  pages = {165104},
  numpages = {22},
  year = {2010},
  month = {Apr},
  publisher = {American Physical Society},
  doi = {10.1103/PhysRevB.81.165104},
  url = {https://link.aps.org/doi/10.1103/PhysRevB.81.165104}
}

@article{corboz2011,
  title = {Stripes in the two-dimensional $t$-${J}$ model with infinite projected entangled-pair states},
  author = {Corboz, Philippe and White, Steven R. and Vidal, Guifr\'e and Troyer, Matthias},
  journal = {Phys. Rev. B},
  volume = {84},
  issue = {4},
  pages = {041108},
  numpages = {5},
  year = {2011},
  month = {Jul},
  publisher = {American Physical Society},
  doi = {10.1103/PhysRevB.84.041108},
  url = {https://link.aps.org/doi/10.1103/PhysRevB.84.041108}
}

@article{liu2017,
  title = {Gradient optimization of finite projected entangled pair states},
  author = {Liu, Wen-Yuan and Dong, Shao-Jun and Han, Yong-Jian and Guo, Guang-Can and He, Lixin},
  journal = {Phys. Rev. B},
  volume = {95},
  issue = {19},
  pages = {195154},
  numpages = {8},
  year = {2017},
  month = {May},
  publisher = {American Physical Society},
  doi = {10.1103/PhysRevB.95.195154},
  url = {https://link.aps.org/doi/10.1103/PhysRevB.95.195154}
}

@article{liu2021,
  title = {Accurate simulation for finite projected entangled pair states in two dimensions},
  author = {Liu, Wen-Yuan and Huang, Yi-Zhen and Gong, Shou-Shu and Gu, Zheng-Cheng},
  journal = {Phys. Rev. B},
  volume = {103},
  issue = {23},
  pages = {235155},
  numpages = {13},
  year = {2021},
  month = {Jun},
  publisher = {American Physical Society},
  doi = {10.1103/PhysRevB.103.235155},
  url = {https://link.aps.org/doi/10.1103/PhysRevB.103.235155}
}

@article{liu2022gapless,
title = {Gapless quantum spin liquid and global phase diagram of the spin-1/2 ${J}_{1}-{J}_{2}$ square antiferromagnetic \text{Heisenberg} model},
author = {Wen-Yuan Liu and Shou-Shu Gong and Yu-Bin Li and Didier Poilblanc and Wei-Qiang Chen and Zheng-Cheng Gu},
journal = {Science Bulletin},
year = 2022,
volume = {67},
pages = {1034-1041},
issn = {2095-9273},
number = {10},
doi = {https://doi.org/10.1016/j.scib.2022.03.010},
url = {https://www.sciencedirect.com/science/article/pii/S2095927322001001},
}

@article{liu2022emergence,
  title = {Emergence of Gapless Quantum Spin Liquid from Deconfined Quantum Critical Point},
  author = {Liu, Wen-Yuan and Hasik, Juraj and Gong, Shou-Shu and Poilblanc, Didier and Chen, Wei-Qiang and Gu, Zheng-Cheng},
  journal = {Phys. Rev. X},
  volume = {12},
  issue = {3},
  pages = {031039},
  numpages = {17},
  year = {2022},
  month = {Sep},
  publisher = {American Physical Society},
  doi = {10.1103/PhysRevX.12.031039},
  url = {https://link.aps.org/doi/10.1103/PhysRevX.12.031039}
}

@article{liu2025hubbard,
  title = {Accurate Simulation of the Hubbard Model with Finite Fermionic Projected Entangled Pair States},
  author = {Liu, Wen-Yuan and Zhai, Huanchen and Peng, Ruojing and Gu, Zheng-Cheng and Chan, Garnet Kin-Lic},
  journal = {Phys. Rev. Lett.},
  volume = {134},
  issue = {25},
  pages = {256502},
  numpages = {8},
  year = {2025},
  month = {Jun},
  publisher = {American Physical Society},
  doi = {10.1103/r4q9-4yvj},
  url = {https://link.aps.org/doi/10.1103/r4q9-4yvj}
}

@article{tocchioBackflowCorrelationsHubbard2011,
	title = {Backflow correlations in the {Hubbard} model: {An} efficient tool for the study of the metal-insulator transition and the large- {U} limit},
	volume = {83},
	copyright = {http://link.aps.org/licenses/aps-default-license},
	issn = {1098-0121, 1550-235X},
	shorttitle = {Backflow correlations in the {Hubbard} model},
	url = {https://link.aps.org/doi/10.1103/PhysRevB.83.195138},
	doi = {10.1103/PhysRevB.83.195138},
	number = {19},
	urldate = {2025-11-02},
	journal = {Phys. Rev. B},
	author = {Tocchio, Luca F. and Becca, Federico and Gros, Claudius},
	month = may,
	year = {2011},
	pages = {195138},
}

@article{cassellaDiscoveringQuantumPhase2023,
	title = {Discovering {Quantum} {Phase} {Transitions} with {Fermionic} {Neural} {Networks}},
	volume = {130},
	issn = {0031-9007, 1079-7114},
	url = {https://link.aps.org/doi/10.1103/PhysRevLett.130.036401},
	doi = {10.1103/PhysRevLett.130.036401},
	number = {3},
	urldate = {2025-11-03},
	journal = {Phys. Rev. Lett.},
	author = {Cassella, Gino and Sutterud, Halvard and Azadi, Sam and Drummond, N. D. and Pfau, David and Spencer, James S. and Foulkes, W. M. C.},
	month = jan,
	year = {2023},
	pages = {036401},
}

@article{cassellaNeuralNetworkVariational2024,
	title = {Neural network variational {Monte} {Carlo} for positronic chemistry},
	volume = {15},
	issn = {2041-1723},
	url = {https://www.nature.com/articles/s41467-024-49290-1},
	doi = {10.1038/s41467-024-49290-1},
	number = {1},
	urldate = {2025-11-03},
	journal = {Nat Commun},
	author = {Cassella, Gino and Foulkes, W. M. C. and Pfau, David and Spencer, James S.},
	month = jun,
	year = {2024},
	pages = {5214},
}

@article{feynmanEnergySpectrumExcitations1956,
	title = {Energy {Spectrum} of the {Excitations} in {Liquid} {Helium}},
	volume = {102},
	issn = {0031-899X},
	url = {https://link.aps.org/doi/10.1103/PhysRev.102.1189},
	doi = {10.1103/PhysRev.102.1189},
	number = {5},
	urldate = {2025-11-02},
	journal = {Phys. Rev.},
	author = {Feynman, R. P. and Cohen, Michael},
	month = jun,
	year = {1956},
	pages = {1189--1204},
}

@article{liFermionicNeuralNetwork2022,
	title = {Fermionic neural network with effective core potential},
	volume = {4},
	issn = {2643-1564},
	url = {https://link.aps.org/doi/10.1103/PhysRevResearch.4.013021},
	doi = {10.1103/PhysRevResearch.4.013021},
	number = {1},
	urldate = {2025-11-03},
	journal = {Phys. Rev. Research},
	author = {Li, Xiang and Fan, Cunwei and Ren, Weiluo and Chen, Ji},
	month = jan,
	year = {2022},
	pages = {013021},
}

@article{qinBenchmarkStudyTwodimensional2016,
	title = {Benchmark study of the two-dimensional {Hubbard} model with auxiliary-field quantum {Monte} {Carlo} method},
	volume = {94},
	issn = {2469-9950, 2469-9969},
	url = {https://link.aps.org/doi/10.1103/PhysRevB.94.085103},
	doi = {10.1103/PhysRevB.94.085103},
	number = {8},
	urldate = {2025-11-03},
	journal = {Phys. Rev. B},
	author = {Qin, Mingpu and Shi, Hao and Zhang, Shiwei},
	month = aug,
	year = {2016},
	pages = {085103},
}

@article{sorella2008backflow,
  title = {Role of backflow correlations for the nonmagnetic phase of the $t\text{--}{t}^{\ensuremath{'}}$ Hubbard model},
  author = {Tocchio, Luca F. and Becca, Federico and Parola, Alberto and Sorella, Sandro},
  journal = {Phys. Rev. B},
  volume = {78},
  issue = {4},
  pages = {041101(R)},
  numpages = {4},
  year = {2008},
  month = {Jul},
  publisher = {American Physical Society},
  doi = {10.1103/PhysRevB.78.041101},
  url = {https://link.aps.org/doi/10.1103/PhysRevB.78.041101}
}

@article{xuCoexistenceSuperconductivityPartially2024,
	title = {Coexistence of superconductivity with partially filled stripes in the hubbard model},
	volume = {384},
	issn = {0036-8075, 1095-9203},
	url = {https://www.science.org/doi/10.1126/science.adh7691},
	doi = {10.1126/science.adh7691},
	number = {6696},
	urldate = {2025-09-08},
	journal = {Science},
	author = {Xu, Hao and Chung, Chia-Min and Qin, Mingpu and Schollwöck, Ulrich and White, Steven R. and Zhang, Shiwei},
	month = may,
	year = {2024},
	pages = {eadh7691},
}

@article{zheng2017stripe,
	title = {Stripe order in the underdoped region of the two-dimensional {Hubbard} model},
	volume = {358},
	issn = {0036-8075, 1095-9203},
	url = {http://arxiv.org/abs/1701.00054},
	doi = {10.1126/science.aam7127},
	number = {6367},
	urldate = {2025-10-14},
	journal = {Science},
	author = {Zheng, Bo-Xiao and Chung, Chia-Min and Corboz, Philippe and Ehlers, Georg and Qin, Ming-Pu and Noack, Reinhard M. and Shi, Hao and White, Steven R. and Zhang, Shiwei and Chan, Garnet Kin-Lic},
	month = dec,
	year = {2017},
	note = {arXiv: 1701.00054 [cond-mat]},
	pages = {1155--1160},
}

@article{zhou2024solving,
	title = {Solving fermi-hubbard-type models by tensor representations of backflow corrections},
	volume = {109},
	issn = {2469-9950, 2469-9969},
	url = {https://link.aps.org/doi/10.1103/PhysRevB.109.245107},
	doi = {10.1103/PhysRevB.109.245107},
	number = {24},
	urldate = {2025-11-02},
	journal = {Phys. Rev. B},
	author = {Zhou, Yu-Tong and Zhou, Zheng-Wei and Liang, Xiao},
	month = jun,
	year = {2024},
	pages = {245107},
}

@article{hermannDeepNeuralNetwork2020a,
	title = {Deep neural network solution of the electronic {Schrödinger} equation},
	volume = {12},
	issn = {1755-4330, 1755-4349},
	url = {http://arxiv.org/abs/1909.08423},
	doi = {10.1038/s41557-020-0544-y},
	number = {10},
	urldate = {2025-11-04},
	journal = {Nat. Chem.},
	author = {Hermann, Jan and Schätzle, Zeno and Noé, Frank},
	month = oct,
	year = {2020},
	pages = {891--897},
}
    
	\begin{widetext}
		\section{Supplemental Material}
		
		\subsection{1. Expressivity of backflow wavefunction}
		The backflow wavefunction originates from the concept of backflow corrections introduced by Feynman and Cohen to describe liquid helium~\cite{feynmanEnergySpectrumExcitations1956}. It incorporates correlations by making the single-particle orbitals depend on the configuration of all other particles, providing a compelling way to improve upon mean-field Slater determinants in fermionic systems~\cite{sorella2008backflow,tocchioBackflowCorrelationsHubbard2011}, which has proven promising in studying various strongly correlated systems~\cite{luo2019backflow,fermionNN2022,clark2024unifying}. Generally, along a chosen site $i$, the amplitude for the configuration $\ket{\bf s}$ of a single-determinant backflow ansatz can be expressed as 
		\begin{equation}
			W(\mathbf{s})=\det(\Psi({\bf s})) = \sum_{m} \psi_m(i,\bar{\mathbf{s}})\, C_m(i,\bar{\mathbf{s}}),
			\label{eq:app_backflow}
		\end{equation}
		where the matrix element $\Psi_{mi}=\psi_m(i,\bar{\mathbf{s}})$ is the backflow-corrected orbital, $ C_m(i,\bar{\mathbf{s}})$ is the corresponding $(m,i)$-cofactor of the matrix $\Psi$, and $m$ labels the orbital which ranges
		from $1$ to particle number $M$. In contrast to the Hartree-Fock approach, where orbitals depend only on the coordinates of a single particle, the backflow wavefunction is much more expressive where each orbital $\psi_m(i,\bar{\mathbf{s}})$ depends on the occupied site $i$ and the full background configuration $\ket{\bar{\mathbf{s}}}$ of the remaining $M-1$ particles. Each pair $(i,\bar{\mathbf{s}})$ uniquely labels a configuration $\mathbf{s}$.

		The Jastrow-Slater wavefunction~\cite{Jastrow1955} is an expressive way to study correlated systems such as fractional quantum Hall effects~\cite{Jain1992jastrow} and electron structures~\cite{hermannDeepNeuralNetwork2020a}. 
		We show that it can also be written as a single-determinant backflow wavefunction. 

		A Jastrow-Slater wavefunction takes the form
		\begin{equation}
			W_{\mathrm{Jastrow}}(\mathbf{s}) = e^{J(\mathbf{s})}\, \det\bigl(\Psi^{\mathrm{HF}}\bigr),
		\end{equation} 
		where $\Psi^{\mathrm{HF}}_{mi} = \psi_m^{\mathrm{HF}}(i)$ are Hartree-Fock orbitals. Expanding the determinant along an arbitrary occupied site $i$ yields  
		\begin{equation}
			W_{\mathrm{Jastrow}}(\mathbf{s}) = e^{J(\mathbf{s})} \sum_{m} \psi_m^{\mathrm{HF}}(i)\, C_m^{\mathrm{HF}}(i,\bar{\mathbf{s}}),
		\end{equation}
		where $C_m^{\mathrm{HF}}$ is the corresponding cofactor. Since $e^{J(\mathbf{s})} = [e^{J(\mathbf{s})/M}]^M$, for an $M$-particle system, we may absorb the Jastrow factor into a redefinition of the orbitals:
		\begin{equation}
			\tilde\psi_m(i,\bar{\mathbf{s}}) = e^{J(\mathbf{s})/M}\,\psi_m^{\mathrm{HF}}(i).
		\end{equation}
		Then the matrix $\tilde\Psi$ with elements $\tilde\Psi_{mi} = \tilde\psi_m(i,\bar{\mathbf{s}})$ satisfies $ W_{\rm Jastrow}(\mathbf{s})=\det(\tilde\Psi) $. Thus in this sense, a Jastrow-Slater wavefunction could be regarded as a special case of the backflow wavefunction.  
		
		\subsection{2. Single-particle-like effective Hamiltonian for the backflow wavefunction}
		
      Here we show how to motivate a single-particle-like eigenvalue problem with an effective Hamiltonian. We consider a fermionic Hubbard-type Hamiltonian 
		\begin{equation}
			H = \sum_{\langle ij \rangle} t_{ij}\, c_i^\dagger c_j + \sum_{ \langle ik \rangle} U_{ik}\, n_i n_k,
		\end{equation} 
		acting on Fock states $|\mathbf{s}\rangle$ with occupation numbers $n_i = c_i^\dagger c_i$. For an $M$-electron system, a general wavefunction is $|\Phi\rangle = \sum_{\mathbf{s}} W(\mathbf{s}) |\mathbf{s}\rangle$. The amplitude of the backflow wavefunction $W(\mathbf{s})$ can be expressed by expanding the determinant along the $i$-th column for any occupied site $i$,
		\begin{equation}
			W(\mathbf{s}) =\det(\Psi) =\sum_{m} \psi_m(i,\bar{\mathbf{s}}) C_m(i,\bar{\mathbf{s}}),
		\end{equation}
		where $\bar{\mathbf{s}}$ (depending on $i$) denotes the background configuration of the remaining electrons after removing the particle at site $i$ from $\mathbf{s}$. The matrix element $\Psi_{mi} = \psi_m(i,\bar{\mathbf{s}})$ is the backflow-corrected orbital, $C_m(i,\bar{\mathbf{s}})$ is the corresponding $(m,i)$-cofactor of the matrix $\Psi$, and $m$ labels the orbitals ($m=1,\dots,M$). In contrast to the Hartree-Fock approach, where
		orbitals depend only on single-particle coordinates, here each orbital $\psi_m(i,\bar{\mathbf{s}})$ depends on the full background configuration of the remaining electrons.
		
		We take the backflow-corrected orbital values $\psi_m(i,\bar{\mathbf{s}})$ as variational parameters. The energy expectation value is $E = \langle\Phi|H|\Phi\rangle/\langle\Phi|\Phi\rangle$. Varying $\psi_m(i,\bar{\mathbf{s}})$ and imposing $\delta E = 0$ then yields the variational condition:
		\begin{equation}
			\langle \frac{\delta\Phi}{\delta\psi_m(i,\bar{\mathbf{s}})} | H - E | \Phi \rangle = 0 .
		\end{equation}
		
		Because $\partial W(\mathbf{s}) / \partial \psi_m(i,\bar{\mathbf{s}}) = C_m(i,\bar{\mathbf{s}})$, the left-hand side of the variational condition can be expanded as
		\begin{equation}
			\langle \frac{\delta\Phi}{\delta\psi_m(i,\bar{\mathbf{s}})} | H - E | \Phi \rangle
			= \sum_{\mathbf{s}'} \frac{\partial W(\mathbf{s}')}{\partial \psi_m(i,\bar{\mathbf{s}})} \langle\mathbf{s}'| H - E |\Phi\rangle
			= C_m(i,\bar{\mathbf{s}}) \langle\mathbf{s}| H - E |\Phi\rangle.
		\end{equation} 
		The pair $(i,\bar{\mathbf{s}})$ determines the full configuration $\mathbf{s}$ uniquely. Hence $\partial W(\mathbf{s}')/\partial \psi_m(i,\bar{\mathbf{s}})$ is nonzero only for $\mathbf{s}' = \mathbf{s}$. Here we implicitly restrict to configurations with $C_m(i, \bar{\mathbf{s}}) \neq 0$, as others give vanishing contributions. 
		
		Consequently, we obtain the equation 
		\begin{equation}
			\langle \mathbf{s}|H|\Phi \rangle = E\langle \mathbf{s}|\Phi \rangle,\qquad \forall\mathbf{s}.
		\end{equation}
		
		Expanding this equation gives
		\begin{equation}
			\begin{aligned}
				\langle \mathbf{s}|H|\Phi \rangle
				&= \sum_{\mathbf{s}'} W(\mathbf{s}')\langle\mathbf{s}|H|\mathbf{s}'\rangle \\
				&= \sum_{ \langle ij \rangle} t_{ij} \sum_{\mathbf{s}'} W(\mathbf{s}') \langle \mathbf{s}|c_i^\dagger c_j|\mathbf{s}'\rangle
				+ \sum_{\langle ik\rangle } U_{ik} \sum_{\mathbf{s}'} W(\mathbf{s}') \langle \mathbf{s}|n_i n_k|\mathbf{s}'\rangle .
			\end{aligned}
		\end{equation}
		
		\begin{figure}
			\centering
			\includegraphics[width=0.5\columnwidth]{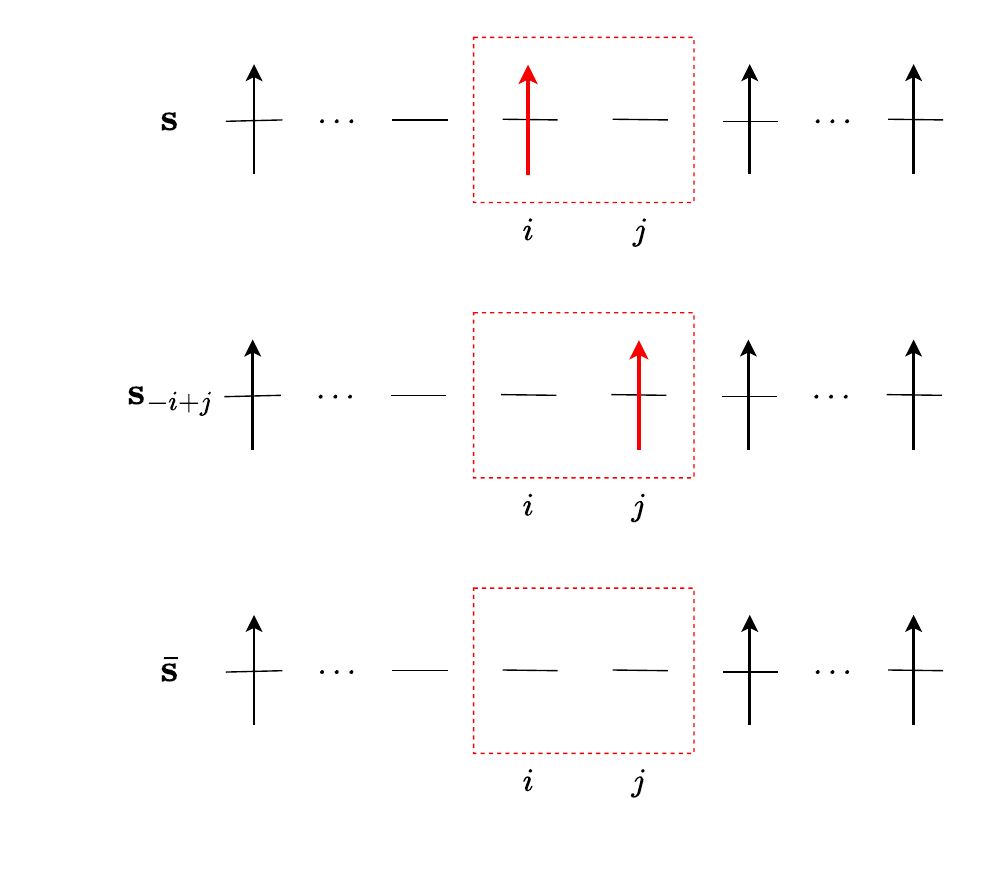}
			\caption{Schematic illustration of $\mathbf{s}$ and $\mathbf{s}_{-i+j}$. The unique configuration $\mathbf{s}_{-i+j}$ is obtained by moving a particle from site $i$ to site $j$. The remaining background configuration $\bar{\mathbf{s}}$ excluding the hopping particle is the same in $\mathbf{s}$ and in $\mathbf{s}_{-i+j}$.}
			\label{fig:tildexillustration}
		\end{figure}
		
		The hopping matrix element is nonzero only when $\mathbf{s}' = \mathbf{s}_{-i+j}$, the configuration obtained by moving one electron from site $i$ to site $j$ in $\mathbf{s}$ (if the move is allowed; otherwise the term vanishes), as illustrated in Figure~\ref{fig:tildexillustration}. We define for the hopping term
		\begin{equation}
			\eta_{ij}^{\mathrm{h}}(\bar{\mathbf{s}}) = \sum_{\mathbf{s}'} \langle \mathbf{s}|c_i^\dagger c_j|\mathbf{s}'\rangle,
		\end{equation}
		which equals the sign factor $\pm 1$ if the move is allowed, and $0$ otherwise. Similarly, for the interaction term
		\begin{equation}
			\eta_{ik}^{\mathrm{I}}(\bar{\mathbf{s}}) = \sum_{\mathbf{s}'} \langle \mathbf{s}|n_i n_k|\mathbf{s}'\rangle.
		\end{equation}
        Both $\eta_{ij}^{\mathrm{h}}$ and $\eta_{ik}^{\mathrm{I}}$ depend only on the configuration $\mathbf{s}$, or equivalently on the pair $(i,\bar{\mathbf{s}})$.
        
		Using the Kronecker delta $\delta_{ij}$ to combine the two sums into a single sum over $i,j$, we obtain
		\begin{equation}
			\langle \mathbf{s}|H|\Phi \rangle = \sum_{ ij } W(\mathbf{s}_{-i+j})\Bigl(t_{ij}\eta_{ij}^{\mathrm{h}}(\bar{\mathbf{s}}) + \delta_{ij}\sum_{k} U_{jk}\eta_{jk}^{\mathrm{I}}(\bar{\mathbf{s}})\Bigr)  =EW(\mathbf{s}).
		\end{equation}
		where the sum over $j$ includes site $i$ and its neighbors (for the hopping terms), and the sum over $k$ includes the neighbors of site $i$ (for the interaction terms).
		
		For an exact eigenstate, the equation $\langle \mathbf{s}|H|\Phi\rangle = E W(\mathbf{s})$ must hold for every $\mathbf{s}$. This condition alone, however, does not
		uniquely determine the individual variational terms. To get a tractable set of equations that can inspire a flexible variational ansatz, we introduce auxiliary quantities $E^{\mathrm{loc}}(i,\bar{\mathbf{s}})$ for each occupied site $i$ and the background configuration $\bar{\mathbf{s}}$, defined as
		\begin{equation}
			E^{\mathrm{loc}}(i,\bar{\mathbf{s}}) = \sum_{j} \frac{W(\mathbf{s}_{-i+j})}{W(\mathbf{s})}\Bigl(t_{ij}\eta_{ij}^{\mathrm{h}}(\bar{\mathbf{s}}) + \delta_{ij}\sum_{k} U_{jk}\eta_{jk}^{\mathrm{I}}(\bar{\mathbf{s}}) \Bigr).\label{eq:local_energy}
		\end{equation}

        This definition closely parallels the concept of the \textit{local energy} $E_{\mathrm{loc}}(\mathbf{s}) = \langle \mathbf{s}|H|\Phi\rangle / W(\mathbf{s})$ commonly used in the scheme of variational Monte Carlo; see Eq.(\ref{eq:vmc}). In VMC, one typically estimates the global energy as the average of $E_{\mathrm{loc}}(\mathbf{s})$ over sampled configurations. 
		Here we decompose the local energy into site-resolved contributions: $E_{\mathrm{loc}}(\mathbf{s}) = \sum_{i} E^{\mathrm{loc}}(i,\bar{\mathbf{s}})$, where the sum runs over all occupied sites $i$ in $\mathbf{s}$. This decomposition is natural because the Hamiltonian consists of single-site (potential) and two-site (hopping) terms, each of which can be associated with specific sites. Since for an exact eigenstate $E_{\mathrm{loc}}(\mathbf{s}) = E$ for all $\mathbf{s}$, summing $E^{\mathrm{loc}}(i,\bar{\mathbf{s}})$ over all occupied sites $i$ in a given configuration $\mathbf{s}$ can yield the total energy $E$.

		Expanding the determinant $W(\mathbf{s}_{-i+j})=\det(\Psi(\mathbf{s}_{-i+j}))$ along the $m$-th row (for a fixed $m$) yields, for each $m$, 
		\begin{equation}
			W(\mathbf{s}_{-i+j}) = \underbrace{\psi_m(j,\bar{\mathbf{s}}) C_m(j,\bar{\mathbf{s}})}_{\displaystyle A_m(j,\bar{\mathbf{s}})} \;+\; 
			\underbrace{\sum_{p\neq j} \psi_m(p,\mathbf{s}^{[p]}) C_m(p,\mathbf{s}^{[p]})}_{\displaystyle B_m(j,\bar{\mathbf{s}})},
			\label{eq:W_expansion}
		\end{equation}
		where $\bar{\mathbf{s}}$ is the background configuration obtained by removing the particle at site $i$ from the original configuration $\mathbf{s}$, and $\mathbf{s}^{[p]}$ denotes the background obtained by removing the particle at site $p$ from $\mathbf{s}_{-i+j}$. Here, $A_m(j,\bar{\mathbf{s}})$ corresponds to the term where the expanded site $j$ has the same background $\bar{\mathbf{s}}$ as the original reference, while $B_m(j,\bar{\mathbf{s}})$ collects all contributions from other sites $p\neq j$ with different backgrounds $\mathbf{s}^{[p]}$.
		
		To isolate the contributions from the $A_m$ terms, we  substitute the full expansion of $W(\mathbf{s}_{-i+j})$ and $W(\mathbf{s})$ from Eq.~\eqref{eq:W_expansion} into Eq.~\eqref{eq:local_energy} and collect all terms that contain $A_m(j,\bar{\mathbf{s}})$ on the left-hand side. Moving all other terms to the right-hand side, we obtain
		\begin{equation}
			\sum_{j}  A_m(j,\bar{\mathbf{s}})\Bigl(t_{ij}\eta_{ij}^{\mathrm{h}}(\bar{\mathbf{s}}) + \delta_{ij}\sum_{k} U_{jk}\eta_{jk}^{\mathrm{I}}(\bar{\mathbf{s}})\Bigr)
			-E^{\mathrm{loc}}(i,\bar{\mathbf{s}}) A_m(i,\bar{\mathbf{s}})
			= R_m(i,\bar{\mathbf{s}}),
		\end{equation}
		where $R_m(i,\bar{\mathbf{s}})$ denotes the remainder that contains all contributions arising from the terms with different background configurations in Eq.~\eqref{eq:W_expansion}. Explicitly,
		\begin{equation}
			R_m(i,\bar{\mathbf{s}})
			= -\!\!\sum_{j} B_m(j,\bar{\mathbf{s}})\Bigl(t_{ij}\eta_{ij}^{\mathrm{h}}(\bar{\mathbf{s}})  + \delta_{ij}\sum_{k} U_{jk}\eta_{jk}^{\mathrm{I}}(\bar{\mathbf{s}}) \Bigr)
			+ E^{\mathrm{loc}}(i,\bar{\mathbf{s}}) B_m(i,\bar{\mathbf{s}}).
		\end{equation}
		
		To obtain an effective single-particle-like description, we define
		\begin{equation}
			\varepsilon_m(i,\bar{\mathbf{s}}) = \frac{R_m(i,\bar{\mathbf{s}})}{\psi_m(i,\bar{\mathbf{s}})\, C_m(i,\bar{\mathbf{s}})},
		\end{equation}
		which formally expresses the original multi-configuration variational condition as a set of single-particle-like equations.
		
        With this we have
		\begin{equation}
			\sum_{j}  \psi_m(j,\bar{\mathbf{s}}) C_m(j,\bar{\mathbf{s}}) \left( t_{ij}\eta^{\mathrm{h}}_{ij}(\bar{\mathbf{s}})  + \delta_{ij}\sum_{k} U_{jk}\eta^{\mathrm{I}}_{jk}(\bar{\mathbf{s}}) \right) = \left[E^{\mathrm{loc}}(i,\bar{\mathbf{s}})+\varepsilon_m(i,\bar{\mathbf{s}})\right]\psi_m(i,\bar{\mathbf{s}}) C_m(i,\bar{\mathbf{s}}).
			\label{eq:app_channel_condition}
		\end{equation}
		
		We perform an additional reparametrization, setting $E^{\mathrm{loc}}(i,\bar{\mathbf{s}})+\varepsilon_m(i,\bar{\mathbf{s}}) = E_m(\bar{\mathbf{s}})\gamma_m(i,\bar{\mathbf{s}})$, where $E_m(\bar{\mathbf{s}})$ plays the role of the eigenvalue for the effective Hamiltonian, and $\gamma_m(i,\bar{\mathbf{s}})$ is a non-zero scaling factor that could be formally absorbed into the Hamiltonian matrix elements. Eventually, we could obtain an effective Hamiltonian:
		\begin{equation}
			\sum_{j} \left[H_m^{\mathrm{eff}}(\bar{\mathbf{s}})\right]_{ij} \psi_m(j,\bar{\mathbf{s}}) = E_m(\bar{\mathbf{s}}) \psi_m(i,\bar{\mathbf{s}}),
			\label{Eq_app:local_eigenproblem}
		\end{equation}
		where the effective Hamiltonian elements read
		\begin{equation}
			\left[H_m^{\mathrm{eff}}(\bar{\mathbf{s}})\right]_{ij} = \frac{1}{\gamma_m(i,\bar{\mathbf{s}})}\frac{C_m(j,\bar{\mathbf{s}})}{ C_m(i,\bar{\mathbf{s}})}\left( t_{ij}\eta^{\mathrm{h}}_{ij}(\bar{\mathbf{s}})  + \delta_{ij}\sum_{k} U_{jk}\eta^{\mathrm{I}}_{jk}(\bar{\mathbf{s}}) \right). 
			\label{Eq_app:H_eff} 
		\end{equation}

        We note the above construction should be viewed as a heuristic derivation for an effective single-particle-like Hamiltonian. The effective Hamiltonian is in general nonlinear and computationally challenging. For each background configuration $\bar{\mathbf{s}}$ and orbital $m$, the effective Hamiltonian $H_m^{\mathrm{eff}}(\bar{\mathbf{s}})$ depends on the cofactor ratio $C_m(j,\bar{\mathbf{s}})/C_m(i,\bar{\mathbf{s}})$, which itself is a function of the orbitals across different backgrounds, and the scaling factor $\gamma_m(i,\bar{\mathbf{s}})$ is also not predetermined. Consequently, solving these equations directly by iteration is impractical. Instead, the effective Hamiltonian motivates the path-expansion ansatz, which we then treat as a variational wavefunction and perform practical calculations within the scheme of variational Monte Carlo.

	\subsection{3. Optimization with variational Monte Carlo}

    We use the variational Monte Carlo (VMC) for optimization. In VMC, the energy function and the $p$-th parameter's gradient are evaluated through the Markov Chain Monte Carlo method:
		\begin{equation}
			\begin{split}
				&	E =\frac{\bra\Phi H \ket \Phi}{\norm\Phi } =\sum_{\mathbf{s}}W(\mathbf{s})^2\frac{\langle{\bf s}| H|\Phi\rangle}{W({\bf s})} =\langle E_{\mathrm{loc}}(\mathbf{s})\rangle ,\\
				&g_p= 2\langle E_{\mathrm{loc}}(\mathbf{s})O_{p}(\mathbf{s}) \rangle
				-2\langle E_{\mathrm{loc}}(\mathbf{s}) \rangle\langle O_{p}(\mathbf{s}) \rangle,
				\label{eq:vmc}
			\end{split}
		\end{equation}
		where the local energy is $E_\mathrm{loc}(\mathbf{s})=\frac{\langle{\bf s}| H |\Phi\rangle}{W({\bf s})} =\sum_{\mathbf{s}'}\frac{W(\mathbf{s}')}{W(\mathbf{s})}\langle \mathbf{s}'| H|\mathbf{s} \rangle$, the $O_{p}(\mathbf{s})=\frac{1}{W(\mathbf{s})}\frac{\partial W(\mathbf{s})}{\partial x_p}$, and $\langle \cdots \rangle$ denotes the average on MCMC samples. $g_p$ is the energy gradients with respect to the $p$-th parameters $x_p$. 
		
		For optimization we employ the stochastic reconfiguration (SR) method~\cite{SR1998, liu2025hubbard},  which is equivalent to the imaginary time evolution using the time-dependent variational principle. SR requires solving the linear system:
		\begin{equation}
			\sum_{q}S_{pq} \dot{x}_q =g_p,
		\end{equation}
		where $S_{pq}=\langle O_p O_q \rangle-\langle O_p \rangle\langle O_q \rangle$. To avoid explicitly constructing the $S$ matrix, we solve the above equation in an iterative way and we only need to know 
		how to map a vector $\dot{x}$ to $y$ under action $S$ to realize $y=S\dot{x}$. Specifically, 
		\begin{equation}
			y_p = \sum_{q} S_{pq} \dot{x}_q = y_p^{(1)} -y_p^{(2)} ,
		\end{equation}
		where 
		\begin{equation}
			\begin{aligned}
				y_p^{(1)} &= \frac{1}{N_{\text{MC}}} \sum_{\mathbf{s}} O_p(\mathbf{s}) \left[ \sum_{q} O_q(\mathbf{s}) \dot{x}_q \right], \\
				y_p^{(2)} &= \langle O_p(\mathbf{s}) \rangle \sum_{q} O_q(\mathbf{s}) \dot{x}_q,
			\end{aligned}
		\end{equation}
		where $N_{\mathrm{MC}}$ is the number of Monte Carlo samples, which is  about $16000$  in our calculations. Note $O_p(\mathbf{s})$ can be stored trivially on different CPU processors,
		while still being convenient to iteratively solve the equation. In practice, we add a small diagonal regularization (e.g., $\epsilon = 10^{-3}$) to $S$ to ensure numerical stability. Once $\dot{x}$ is obtained, parameters are updated as
		\begin{equation}
			x_p(\tau+1) =x_p(\tau) -d\tau \cdot \dot{x}_p ,
		\end{equation}
		where $d\tau$ is the step size, which can be tuned from $d\tau=0.1$ to $0.01$.

    For the residual hierarchical backflow (RHB) wavefunction using $N_{\det}$ determinants, its amplitude $W(\mathbf{s})$   is obtained according to 
	\begin{equation}
		W(\mathbf{s})=\sum_{\alpha=1}^{N_{\det}} \det(\Psi^{[\alpha]}),
	\end{equation}
	with matrix elements
	\begin{equation}
		\Psi^{[\alpha]}_{mi} = \psi_m^{\mathrm{HB}}(i, \bar{\mathbf{s}}) Q_m^{[\alpha]}(\bar{\mathbf{s}}).
	\end{equation}
	All determinant share the same orbitals $\psi_m^{\mathrm{HB}}$ but each carries its own configuration-dependent factor $Q_m^{[\alpha]}(\bar{\mathbf{s}})$ that can be parameterized by a neural network. This multi‑determinant construction, with independent $Q_m^{[\alpha]}(\bar{\mathbf{s}})$ factors, lifts the single‑determinant constraint and provides additional variational freedom beyond the HB backbone, thereby enhancing the expressive power.

			\begin{figure}[htp]
			\centering
			\begin{minipage}[b]{0.48\textwidth}
				\centering
				\begin{overpic}[width=1\textwidth]{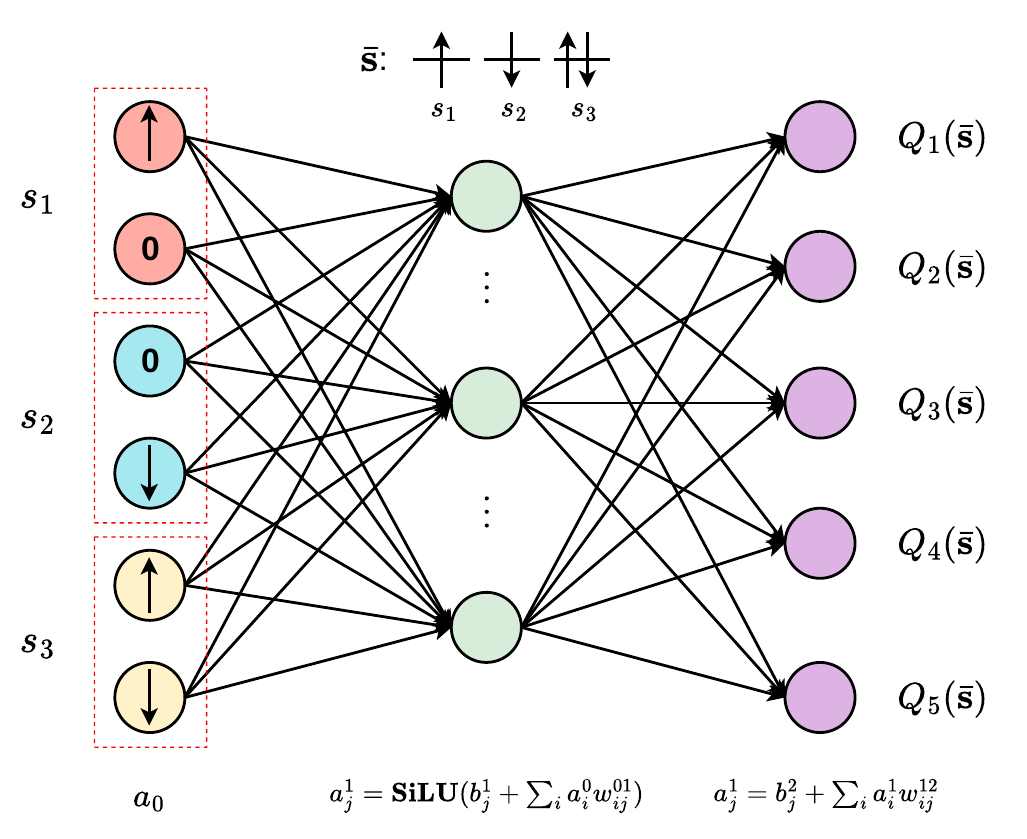}
					\put(5, 80){(a) }  
				\end{overpic}
			\end{minipage}
			\hspace{1pt}
			\begin{minipage}[b]{0.48\textwidth}
				\centering
				\begin{overpic}[width=1\textwidth]{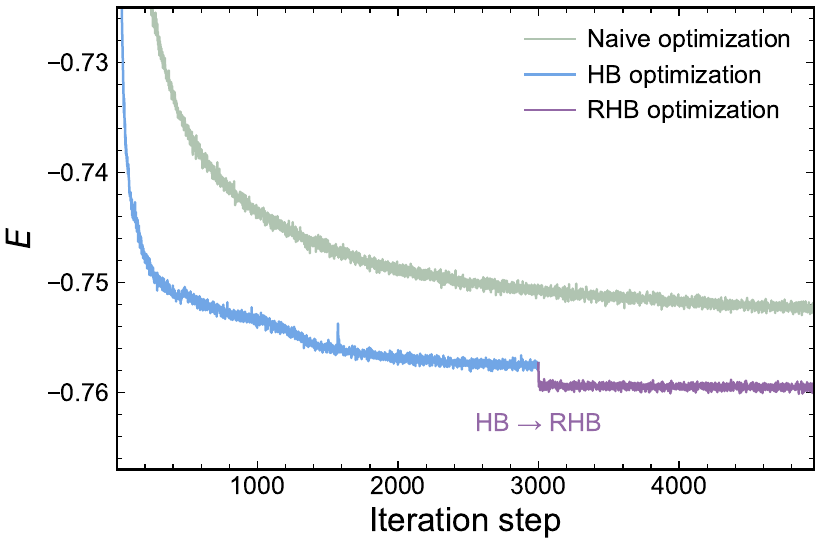}
					\put(5, 80){(b) }  
				\end{overpic}
			\end{minipage}
			\hspace{1pt}
            
			\caption{(a) The structure of the neural network for a single determinant. The input layer has $2N$ neurons ($N$ is the system size), which encode the occupation of the environment $\bar{\mathbf{s}}$ for each site. The output layer specifies the nonlocal factors $Q_m^{[\alpha]}(\bar{\mathbf{s}})$ for each orbital $m$ and determinant $\alpha$, with total $MN_{\rm det}$ outputting values ($M$ is the electron number). (b) Comparison of naive versus two-stage optimization for the same RHB ansatz on a $4\times16$ Hubbard model at $U=8$ and $n_h=0.125$. Both optimizations start from the same Hartree-Fock state. The light green curve shows the naive optimization (single-stage direct optimization), which becomes trapped at a higher energy. The two-stage scheme first optimizes the HB backbone alone, then jointly optimizes both the HB backbone and the nonlocal part.}
			\label{fig:NN_structure_optimization}
		\end{figure}

The hierarchical part  $\psi_m^{\mathrm{HB}}(i, \bar{\mathbf{s}})$ is directly computed as in the HB wavefunction. For the nonlocal part $Q_m^{[\alpha]}(\bar{\mathbf{s}})$, for simplicity we adopt a three-layer fully connected feed-forward neural network (FNN) for the spinful case, as illustrated in Figure~\ref{fig:NN_structure_optimization}(a). The network is described as follow:
	\begin{enumerate}
		
		\item Input Layer: The input layer consists of $2N$ neurons encoding the configuration $\ket{\bar{\mathbf{s}}}$, which is obtained from $\ket{\mathbf{s}}$ by removing the electron at site $i$. For each site $i$, two neurons $(i_a, i_b)$ represent the spin-up and spin-down components, respectively. Neuron $i_a$ outputs $+1$ if site $i$ is occupied by a spin-up electron and $-1$ otherwise; neuron $i_b$ outputs $+1$ if site $i$ is occupied by a spin-down electron and $-1$ otherwise.
		
		\item Hidden Layer: The hidden layer contains $N_{\text{neuron}}$ neurons with the Sigmoid Linear Unit (SiLU) activation function, $\mathrm{SiLU}(x)=x\cdot\mathrm{sigmoid}(x)=x/(1+e^{-x})$. 
		
		\item Output Layer: The output layer contains $N_{\text{det}} \times M$ neurons, each with a linear activation function, outputting the values of the nonlocal factors $Q_m^{[\alpha]}(\bar{\mathbf{s}})$ for each orbital $m$ and determinant $\alpha$. 
		
	\end{enumerate}

	The neural network introduces variational parameters from the two fully connected layers, totaling approximately $4N_{\text{neuron}}N + 2N_{\text{det}} N_{\text{neuron}} M$ (including both weights and biases). The HB part contributes $z d^2 K M N$ parameters. Hence, the total number of variational parameters in the RHB wavefunction scales as $O(N_{\text{neuron}}N) + O(N_{\text{neuron}}M) + O(MN)$. In this work, we typically set $N_{\text{neuron}} = 100$ and $N_{\text{det}} = 5$ (see later).

   The local-nonlocal structure of the RHB enables a two-stage optimization. Concretely, we first optimize the HB backbone alone (take $K=2$), then introduce the nonlocal factor $Q_m^{[\alpha]}(\bar{\mathbf{s}})$ and perform a joint optimization of both the HB and the nonlocal part. Figure~\ref{fig:NN_structure_optimization}(b) compares the optimization behavior using different schemes for the $4\times16$ Hubbard model at $U=8$ and hole doping $n_h=0.125$. In the naive optimization, the  RHB ansatz (including the FNN) is optimized directly from the Hartree‑Fock starting point. The energy converges slowly and becomes trapped at a relatively high value. In contrast, using the two‑stage scheme enabled by the local‑nonlocal structure, we first optimize the HB backbone; this converges rapidly and already reaches an energy lower than that of the naive full optimization. We then add the FNN and continue joint optimization, which further lowers the energy significantly below the naive result. This demonstrates the advantage of the local‑nonlocal structure of RHB for efficient and accurate simulations.

    \subsection{4. Additional Results}
            
    Table~\ref{table:Hubbard_energies_for_n=1} presents energy comparisons for cases of half filling  on the square lattice under PBC. The reference energies are
	from auxiliary-field quantum Monte Carlo (AFQMC) without sign problems~\cite{qinBenchmarkStudyTwodimensional2016}.
	From the
	table, the $K=1$ wavefunction reproduces the AFQMC energies with remarkable accuracy, achieving relative errors on the order of $10^{-3}$ across all cases.
        \begin{table}[h]
		\centering
		\caption{Energies per site for the Hubbard model at half-filling under periodic boundary conditions, for $U = 2$, $4$, $6$, and $8$. For $U = 8$, results are shown for HB depths $K = 0$, $1$, and $2$. The energy sampling errors are around 0.0003. Reference energies are from AFQMC ~\cite{qinBenchmarkStudyTwodimensional2016}.}
		\label{table:Hubbard_energies_for_n=1}
	\begin{tabular}{c| c c | c c |c c |c c c c }
		\hline\hline
	\multirow{2}{*}{Size} & \multicolumn{2}{c|}{$U=2$} & \multicolumn{2}{c|}{$U=4$} & \multicolumn{2}{c|}{$U=6$} & \multicolumn{4}{c}{$U=8$} \\
	\cline{2-3} \cline{4-5} \cline{6-7} \cline{8-11} 
		& $K=1$ & QMC & $K=1$ & QMC & $K=1$ & QMC & $K=0$ (HF) &$K=1$ & $K=2$& QMC \\
		\hline
		$4\times4$   & $-1.1248$ & $-1.1265(4)$ & $-0.8486$ & $-0.8510(4)$ & $-0.6577$ & $-0.6588(4)$ & $-0.4898$ & $-0.5281$ & $-0.5291$ & $-0.5298(1)$\\
		$6\times6$   & $-1.1506$ & $-1.1515(9)$ & $-0.8554$ & $-0.8573(9)$& $-0.5255$ & $-0.5278(3)$ & $-0.4797$ & $-0.5255$ & $-0.5266$ &$-0.5278(3)$ \\
		$8\times8$   & $-1.1626$ & $-1.1636(0)$ & $-0.8581$ & $-0.8601(6)$ & $-0.6554$ & $-0.6587(5)$ & $-0.4743$ & $-0.5229$ & $-0.5245$ & $-0.5262(5)$ \\
		$10\times10$ & $-1.1677$ & $-1.1690(8)$ & $-0.8585$ & $-0.8612(4)$  & $-0.6550$ & $-0.6580(2)$ & $-0.4690$ & $-0.5226$ & $-0.5240$& $-0.5254(3)$\\
		\hline\hline
	\end{tabular}
		\end{table}
 
        Table~\ref{table:4x4} shows the ground-state energy per site for the $4\times4$ Hubbard model at $n_h=0.125$, $U=8$. The HB $K=2$ energy is $-0.7385$. With the RHB ansatz using $N_{\mathrm{det}}=5$ determinants and $N_{\mathrm{neuron}}=100$, the energy improves to $-0.7403$, corresponding to a relative error of $2\times10^{-3}$ compared to the exact diagonalization (ED) result $-0.7418$. Increasing $N_{\mathrm{det}}$ to $16$ (with $N_{\mathrm{neuron}}=80$) further reduces the error to $1.1\times10^{-3}$, but at a higher computational cost. Based on this tradeoff, we adopt $N_{\mathrm{det}}=5$ and $N_{\mathrm{neuron}}=100$ for all subsequent large-scale simulations.
		   	\begin{table*}[h]
			\caption{Energy per site for the $4\times 4$ Hubbard model at filling $n_h=0.125$ and $U=8$ under periodic boundary conditions. The RHB consists of the HB $K=2$ part and a FNN part. ED is the exact diagonalization result.}
			\begin{tabular}{c|cccccccc|c}
				\hline\hline
				Method & $K=1$ & $K=2$ & $K=3$& \multicolumn{4}{c}{RHB ($N_{\mathrm{det}}=5$)} & RHB ($N_{\mathrm{det}}=16$) & ED \\
				\hline
				$N_{\mathrm{neuron}}$ & -- & -- & -- & 16 & 80 & 100 & 160 & 80 & -- \\
				\hline
				$E$ & $-0.7370$ & $-0.7385$& $-0.7390$ & $-0.7395$ & $-0.7395$ & $-0.7403$ & $-0.7401$ & $-0.7410$ & $-0.7418$ \\
				\hline\hline
			\end{tabular}
			\label{table:4x4}
		\end{table*}

    Table~\ref{table: 1/8 doped PBC} compares the ground-state energies for $4\times L$ rectangular lattices at hole doping $n_h=0.125$ and $U=8$ under PBC. Reference energies are from 
	NNB~\cite{luo2019backflow} and HFDS~\cite{fermionNN2022}. NNB replaces the static set of orbitals with a configuration dependent set which is generated by a neural network. HFDS instead works in the paradigm of projected hidden fermions using neural networks to replace the standard Slater Determinant with a larger determinant which includes single-particle orbitals. The two methods have intrinsic relations~\cite{clark2024unifying}.
    		\begin{table}[h]
			\centering
			\caption{Energies persite for the Hubbard model at $n_h = 0.125$, $U = 8$  for different systems under periodic boundary conditions. The energy sampling errors are around 0.0003. Energies of HFDS~\cite{fermionNN2022} and NNB~\cite{luo2019backflow} are listed for comparison.}
			\label{table: 1/8 doped PBC}
			\begin{tabular}{c c c c c c c}
				\hline\hline
				size & $K=0$ & $K=1$ & $K=2$ & RHB & HFDS  & NNB \\
				\hline
				$4\times4$  &$-0.6272$ & $-0.7370$ & $-0.7385$ & $-0.7410$ & $-0.7409$ &  $-0.730$ \\
				$4\times8$ & $-0.6146$ & $-0.7599$ & $-0.7617$ & $-0.7641$ & $-0.7633$ &  $-0.755$ \\
				$4\times12$& $-0.6058$ & $-0.7598$ & $-0.7613$ & $-0.7630$ & -- &  $-0.746$ \\
				$4\times16$&$-0.6035$ & $-0.7585$ & $-0.7597$ & $-0.7611$ & $-0.7530$ &  $-0.746$ \\
				\hline\hline
			\end{tabular}
		\end{table}

     We also compare the local spin moment $\langle S^z \rangle$ and the hole density distributions for the $4\times16$ lattice at $n_h=0.125$. Figure~\ref{fig:fig4x16pbc} shows that the distributions obtained with $K=1$, $K=2$, and the RHB are nearly identical. This indicates that the essential spatial structure of spin and charge is already captured by the hierarchical backflow component $\psi_m^{\mathrm{HB}}$, and the nonlocal factor $Q_m^{[\alpha]}(\bar{\mathbf{s}})$ in this case primarily provides an energy refinement. 
		\begin{figure}[htp]
			\centering
			\includegraphics[width=1\textwidth]{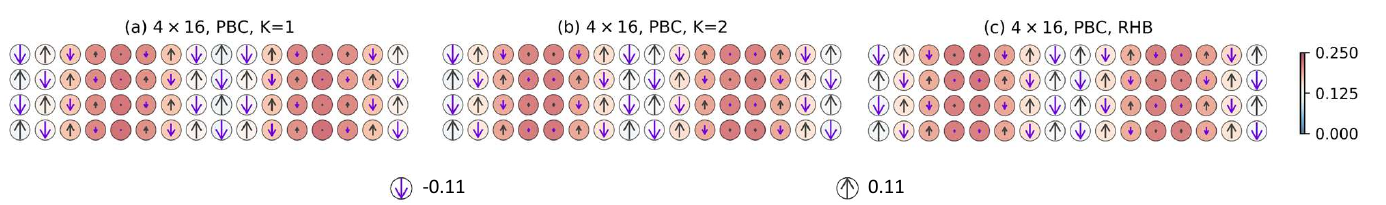}
			\caption{Spatial distribution of hole density (color map) and spin moment along the $z$-axis (arrows) for the Hubbard model under periodic boundary conditions. (a-c) Results for a $4\times 16$ lattice with hole density $n_h=0.125$, obtained using three different methods: (a) $K=1$, (b) $K=2$, and (c) RHB.}
			\label{fig:fig4x16pbc}
		\end{figure}

	\end{widetext}

\end{document}